\documentclass[amsmath,twoside,amsfonts,superscriptaddress, reprint, prx]{revtex4-2}

\usepackage{amssymb,graphicx,physics,mhchem,xcolor}
\allowdisplaybreaks

\makeatletter
\g@addto@macro\bfseries{\boldmath}
\makeatother

\usepackage[colorlinks=true, linkcolor=blue, citecolor=blue,
urlcolor=blue, bookmarks=true, hypertexnames=true]{hyperref}

\begin{document}	

\title{Spectroscopic Evidence of Competing Diagonal Spin Interactions and Spin Disproportionation 
in the Bilayer Nickelate La$_3$Ni$_2$O$_7$}

\date{submitted 5 February 2026;\enspace revised 12 September 2026}

\begin{abstract}		
A comprehensive spectroscopic map of the electronic, magnetic, and lattice excitations is presented for the bilayer nickelate La$_3$Ni$_2$O$_7$    using Raman scattering at ambient pressure. 
Upon entering the spin density wave state below 153~K, the $A_{1g}$ channel exhibits an abrupt electronic spectral gap with a clear isosbestic point. 
In contrast, the $B_{1g}$ and $B_{2g}$ channels are dominated by pronounced two-magnon (2M) excitations, representing a signature of incipient Mottness. 
These 2M signals in both channels constitute evidence for two distinct in-plane spin exchange interactions along the Ni--O bonding and its diagonal directions. 
Calculations based on the spin wave theory further reveal that the 2M mode in the $B_{2g}$ channel arises from the competition between two bond-diagonal antiferromagnetic interactions mediated by nickel $d_{x^2-y^2}$ orbitals. 
Furthermore, emergent low-energy 2M excitations below 10~meV are found to originate from distinct, weaker spin moments,  supporting spin disproportionation. 
Simultaneously, an anomalous softening of $B_{1g}$ phonons from 280 down to 4.5~K is uncovered, suggesting the presence of an incipient lattice instability leading to checkerboard-type breathing modulations. 
Collectively, these findings identify a ground state of the bilayer nickelate characterized by competing bond-diagonal interactions, spin disproportionation, and an incipient lattice instability, establishing key ingredients for understanding the mechanism of nickelate superconductivity.
\end{abstract}

\author{Dong--Hyeon Gim}
\affiliation{Department of Physics and Astronomy, Seoul National University, Seoul 08826, Korea}

\author{Dirk Wulferding}
\affiliation{Department of Physics and Astronomy, Sejong University, Seoul 05006, Korea}

\author{Hengyuan Zhang}
\author{Deyuan Hu}
\author{Meng Wang}
\email{wangmeng5@mail.sysu.edu.cn}
\affiliation{Institute of Neutron Science and Technology, Guangdong Provincial Key Laboratory of Magnetoelectric Physics and Devices, School of Physics, Sun Yat-Sen University, Guangzhou, Guangdong 510275, China}

\author{Kee~Hoon~Kim}
\email{optopia@snu.ac.kr}
\affiliation{Department of Physics and Astronomy, Seoul National University, Seoul 08826, Korea}
\affiliation{Institute of Applied Physics, Seoul National University, Seoul 08826, Korea \vspace{0.8em}}

\maketitle		

\section{Introduction} \label{sec:1}	
The discovery of pressure-induced high-temperature superconductivity (SC) in the bilayer nickelate \ce{La3Ni2O7}\  with a critical temperature as high as $T_c = 80$~K \cite{sun23nature} has sparked intense interest in the mechanisms of unconventional SC. 
Given that \ce{La3Ni2O7}\ exhibits a spin density wave (SDW) transition at lower pressures \cite{chen24prl, plokhikh25arxiv}, the interplay between the SDW and SC is expected to be crucial, which has motivated numerous theoretical proposals explaining the intertwined ground states based on the SDW and SC. 
However, the nature of SC in \ce{La3Ni2O7}\ remains a subject of intense debate. Theories based on various models have led to divergent scenarios on the possible pairing mechanism \cite{liu23prl, sakakibara24prl, zhang24nc, ryee24prl, qu24prl, lu24prl, jiang24prl, yang23prb, lechermann23prb, fan24prb, heier24prb, chen24prb, bejas25prb, xi25prb, liu25prb, braz25prr, luo24npjqm}. 
Experimentally, investigations on the pressurized crystals and strained films have resulted in conflicting interpretations regarding whether the primary SC gap symmetry is a fully gapped $s$-wave or a nodal $d$-wave \cite{guo25nc, cao25arxiv, fan25arxiv}. 
Given that the emergence of SC is intimately linked to the underlying electronic and magnetic correlations, understanding the parent SDW state provides an important foundation for elucidating the SC mechanism.

Nevertheless, beyond its ordering wave vector $\mathbf{Q}_\mathrm{SDW}=(\pi/2,\pi/2)$ \cite{plokhikh25arxiv}, the exact spin configuration of the SDW state remains controversial.
Although the magnon dispersion measurements by resonant inelastic x-ray scattering (RIXS) have revealed sizable interlayer ($J_c$) and secondary intralayer ($J_3$) spin interactions \cite{chen24nc}, the resulting spin pattern---whether spin-charge or double-spin stripes---is undetermined.
An inelastic neutron scattering (INS) study could not detect signatures of a long-range spin ordering, presumably because magnetic moments were too weak \cite{xie24sb}. 
Meanwhile, the muon spin rotation and relaxation ($\mu$SR) and neutron powder diffraction (NPD) studies suggested the coexistence of two nickel spin moments ($m_1 = 0.66$ and $m_2 = 0.05\mu_\mathrm{B}$) \cite{plokhikh25arxiv, khasanov25np}. 
Two x-ray absorption spectroscopy (XAS) and x-ray scattering studies on \ce{La3Ni2O7}\ films concluded that nickel spins are aligned in the $ab$-plane, being perpendicular to $\mathbf{Q}_\mathrm{SDW}$, but suggesting different spin-stripe patterns  \cite{ren25cp, gupta25nc}.
Moreover, a nuclear magnetic resonance study suggested rather  small but uniform out-of-plane spin moments of $\sim0.1\ensuremath{\mu_\mathrm{B}}$, supporting the double-spin stripe pattern \cite{zhao25sb}. 
On the other hand, a nuclear quadrupolar resonance study raised the possibility of coexisting charge density wave (CDW) states, adding another layer of complexity \cite{luo25cpl}.

\begin{figure}[] 		
{\includegraphics[width=0.85\columnwidth,draft=false]{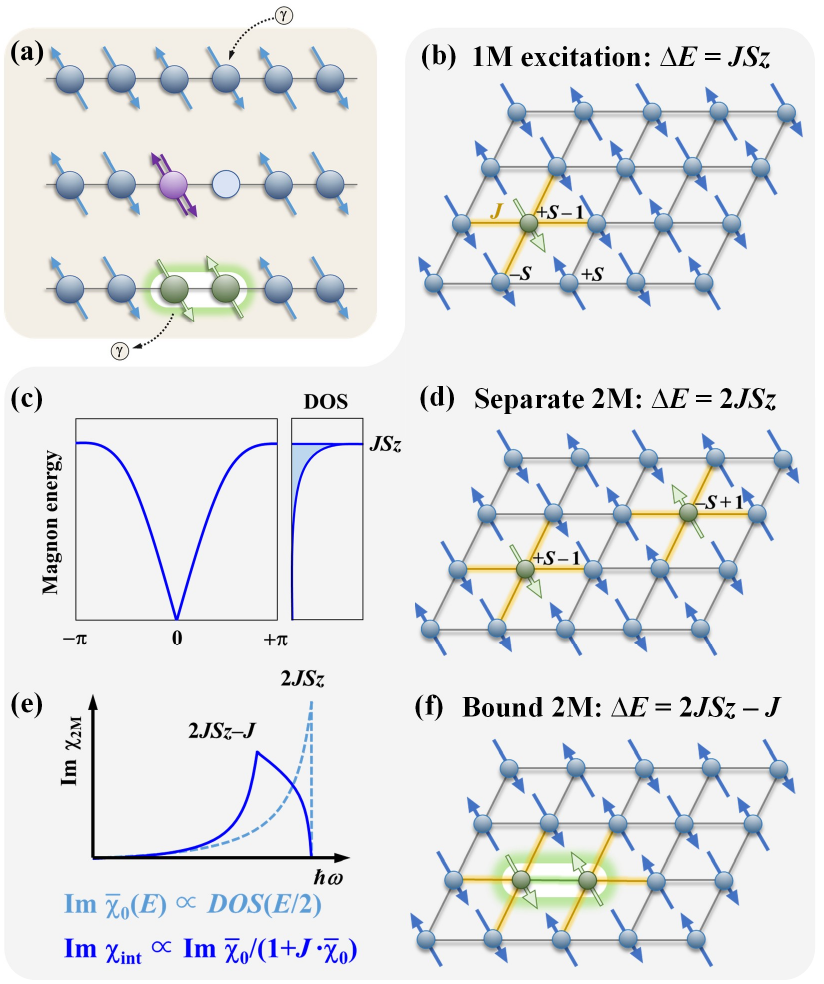}}  \vspace{-.8em}
\caption{\label{fig1}%
Raman scattering process of 2M excitations. (a) The Fleury--Loudon exchange-scattering mechanism in antiferromagnets. An incident photon ($\gamma$) virtually excites an electron to a neighboring site; the subsequent exchange process creates two adjacent, opposite spin-flips in the final state. (b)-(f) Energy costs of spin-flips. $z$ is the number of nearest neighbors. Single 1M spin-flip (b) spectrum peaks at $JSz$ [(c), right panel]. Spectrum of a distant, non-interacting 2M excitation (d) peaks at $2JSz$ [(e), dashed line]. Renormalized spectrum of an adjacent, interacting 2M excitation (f) peaks at a lower energy shifted by $J$ [(e), solid line].}
\end{figure}

The conflicting interpretations from various experiments have precluded a coherent understanding of the electronic and magnetic properties of \ce{La3Ni2O7}. 
To elucidate the essential microscopic interactions in \ce{La3Ni2O7}, we employ comprehensive, symmetry-resolved Raman spectroscopy (see Appendix~\ref{app:a} for experimental details). 
Raman scattering is uniquely sensitive to antiferromagnetic (AF) spin-flip processes, specifically two-magnon (2M) excitations (Fig.~\ref{fig1}), which are governed by the Fleury--Loudon (FL) exchange scattering mechanism \cite{fleury68pr}. 
The corresponding scattering Hamiltonian, which depends on the incident ($\mathbf{\hat{e}}_\mathrm{I}$) and scattered ($\mathbf{\hat{e}}_\mathrm{S}$) light polarizations, is given by:
\begin{equation}	\label{eq1}	
\mathcal{H}_\mathrm{FL}=\sum_{\langle i,j \rangle} J_{ij} 
(\mathbf{\hat{e}}_\mathrm{I}\cdot\mathbf{\hat{r}}_{ij})
(\mathbf{\hat{e}}_\mathrm{S}\cdot\mathbf{\hat{r}}_{ij})
\mathbf{S}_i \cdot \mathbf{S}_j,
\end{equation}
where $J_{ij}$ is the exchange interaction between spins $\mathbf{S}_{i}$ and $\mathbf{S}_{j}$ on sites $i$ and $j$, and $\mathbf{\hat{r}}_{ij}$ is the unit vector connecting them.
As Eq.~(\ref{eq1}) demonstrates, the Raman selection rules establish an important link between the polarization configuration ($\mathbf{\hat{e}}_\mathrm{I}\mathbf{\hat{e}}_\mathrm{S}$) and the geometry of magnetic bonds. 
Consequently, the 2M Raman response provides energy- and symmetry-dependent constraints on the spin interactions in antiferromagnets \cite{fleury68prl, fleury70prl, brya74prb, brya73aipcp, lockwood87prb, singh89prl, sulewski91prl, choi08prb, li20prl, sugai03prb, onose04prb}, which are crucial for stabilizing the magnetic ground states.

\section{Results} \label{sec:2}	
\begin{figure*}[t!] 		
{\includegraphics[width=1.75\columnwidth,draft=false]{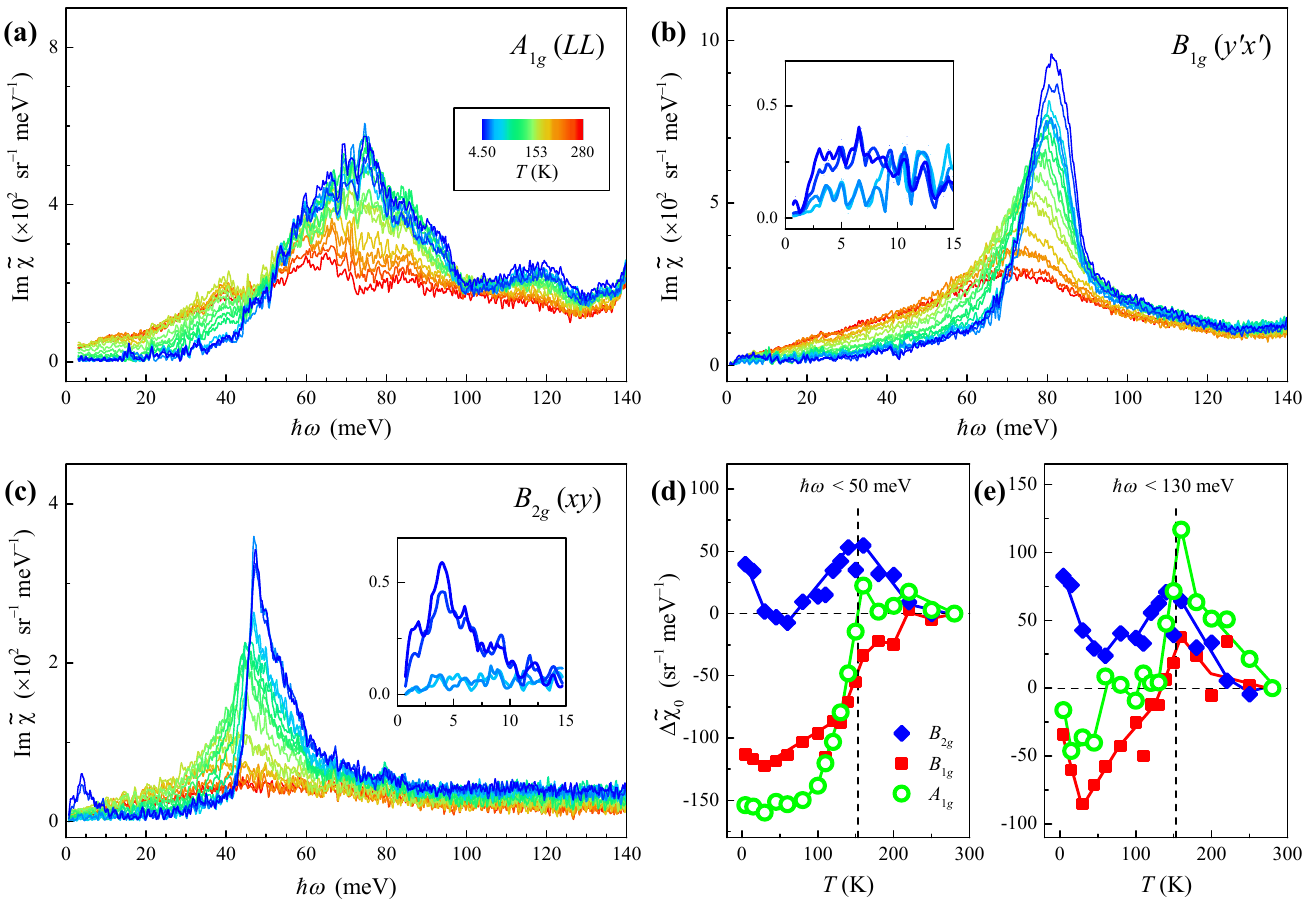}} \vspace{-.8em}
\caption{\label{fig2}%
Temperature-dependent Raman response of \ce{La3Ni2O7}. (a)-(c) Phonon-subtracted Raman response in the \ensuremath{A_{1g}}\ $(LL)$, \ensuremath{B_{1g}}\ $(y'x')$, and \ensuremath{B_{2g}}\ $(xy)$ symmetry channels, respectively. Insets in (b),(c) show the low-energy region at 4.5, 15, 30, and 45~K. (d),(e) SW difference $\Delta\tilde{\chi}_0$ in each scattering channel is integrated up to 50 and 130~meV, respectively. Solid lines are guides for the eye. Vertical dashed lines represent \ensuremath{T_\mathrm{SDW}}.
}
\end{figure*}

Figures~\ref{fig2}(a)-(c) present the phonon-subtracted \ce{La3Ni2O7}\ Raman response ($\Im \tilde\chi$) measured in three polarization configurations: $(LL)$, $(y'x')$, and $(xy)$. 
Here, $L$ denotes left circularly polarized light, $x$ and $y$ represent linear polarizations parallel to the in-plane tetragonal axes, and $x'$ and $y'$ are rotated by 45\textdegree\ (details on crystal orientation and phonon subtraction are provided in Appendices~\ref{app:a} and \ref{app:b}). 
Within the quasi-tetragonal $D_{4h}$ description, these scattering geometries selectively probe excitations in the \ensuremath{A_{1g}}\ ($s$-wave), \ensuremath{B_{1g}}\ (\ensuremath{d_{x^2-y^2}}-wave), and \ensuremath{B_{2g}}\ ($d_{xy}$-wave) symmetry channels, respectively (Appendix~\ref{app:c}). Across all three symmetry channels, the \ce{La3Ni2O7}\ Raman response exhibits a pronounced temperature evolution depending on the energy $\hslash\omega$,
significantly differing from those of the single-layer ({\ce{La2NiO4}}) and trilayer (\ce{La4Ni3O10}) nickelates \cite{sugai90prb, sugai98jpsj, gim25prl}.

Upon cooling below the SDW transition temperature \ensuremath{T_\mathrm{SDW}}\ = 153~K, the \ensuremath{A_{1g}}\ channel exhibits a clear signature of an electronic gap opening with the resulting isosbestic point near 50~meV [Fig.~\ref{fig2}(a)]:
While the low-energy spectral weight (SW) below 50~meV remains nearly constant above \ensuremath{T_\mathrm{SDW}}, it is sharply suppressed below \ensuremath{T_\mathrm{SDW}}, depleted to a value close to zero. 
The low-energy dynamic susceptibility difference 
$\Delta\tilde\chi_0(T)=\int_\mathrm{3~meV}^\mathrm{50~meV}d\hslash\omega \Im [\tilde\chi(T)-\tilde\chi(\mathrm{280~K})]/\hslash\omega$
integrated up to the isosbestic point also decreases sharply below \ensuremath{T_\mathrm{SDW}} [green circles in Fig.~\ref{fig2}(d)], corroborating the gap opening behavior. 
Notably, the electronic SDW gap in the trilayer nickelate \ce{La4Ni3O10}\ also develops most prominently in the \ensuremath{A_{1g}}\ Raman channel \cite{gim25prl}. 
This suggests that the SDW formation in both bilayer and trilayer nickelates may involve similar electronic reconstructions, leaving a fingerprint in the four-fold symmetric scattering channel.

In contrast, the SWs in the \ensuremath{B_{1g}}\ and \ensuremath{B_{2g}}\ channels show complex behaviors starting well above \ensuremath{T_\mathrm{SDW}}\ [Figs.~\ref{fig2}(d),(e)], driven by the high-energy peaks located above 40~meV (see Appendix~\ref{app:d} for detailed analysis). Furthermore, as $T$ drops below 30~K, distinct low-energy humps emerge below 15~meV in both the \ensuremath{B_{1g}}\ and \ensuremath{B_{2g}}\ channels [insets of Figs.~\ref{fig2}(b),(c)].

\section{Multichannel spin-exchange excitations}\label{sec:3}  
The prominent peaks in the \ensuremath{B_{1g}}\ and \ensuremath{B_{2g}}\ channels are characteristic signatures of 2M excitations generated by two opposite spin-flips at adjacent sites (Fig.~\ref{fig1}; Appendix~\ref{app:d}). 
The simultaneous observation of the prominent 2M peaks in both \ensuremath{B_{1g}}\ and \ensuremath{B_{2g}}\ channels marks a key feature distinguishing \ce{La3Ni2O7}\ from cuprate superconductors. 
In cuprates, the nearest-neighbor exchange interaction $J_1$ is dominant, generating 2M peaks near $3.6 J_1$ exclusively in the \ensuremath{B_{1g}}\ channel \cite{singh89prl, sulewski91prl}. 
In contrast, the magnetic \ensuremath{B_{2g}}\ response in cuprates appears only as a much broader and weaker feature at a higher energy, originating from higher-order processes involving itinerant charge carriers or quantum fluctuations \cite{singh89prl, sugai13jpcm}.

\begin{figure}[] 		
{\includegraphics[width=0.8\columnwidth,draft=false]{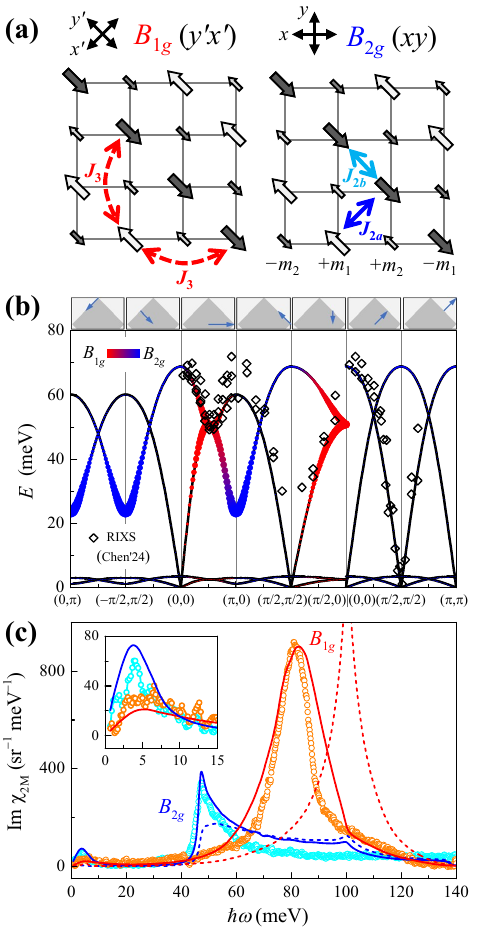}} \vspace{-.8em}
\caption{\label{fig3}%
2M excitations and multichannel exchange interactions. (a) Schematic of the magnetic structure and exchange interactions probed in each symmetry channel. (b) Calculated magnon dispersion relations (lines) using parameters from Table~\ref{tab1}, compared with the RIXS data (squares) \cite{chen24nc}. Color scale indicates the calculated strength of Raman scattering matrix elements for the \ensuremath{B_{1g}}\ (red) and \ensuremath{B_{2g}}\ (blue) channels. (c) Calculated bare (dashed) and renormalized (solid lines) 2M response in the \ensuremath{B_{1g}}\ (red) and \ensuremath{B_{2g}}\ (blue) channels are compared with the experimental data at 4.5~K (orange and cyan circles). For this comparison, we subtracted a small high-energy background, assuming quadratic energy dependence, from the \ensuremath{B_{1g}}\ data [Fig.~\ref{fig2}(b)]. Inset: a magnified view below 15 meV, showing features arising from the low-moment ($\pm m_2$) sublattices.}
\end{figure}

In sharp contrast, in \ce{La3Ni2O7}\, the \ensuremath{B_{2g}}\ signal is detected as a sharp peak with substantial intensity (approximately 37\% of the \ensuremath{B_{1g}}\ intensity), and its peak energy (47~meV at 4.5~K) is even lower than that of the \ensuremath{B_{1g}}\ peak (81~meV). 
These observations  indicate that the \ensuremath{B_{1g}}\ and \ensuremath{B_{2g}}\ peaks in \ce{La3Ni2O7}\ arise from the same fundamental mechanism, i.e., the FL exchange scattering process of 2M excitations. 
As evident from Eq.~(\ref{eq1}), the 2M signal in the \ensuremath{B_{1g}}\ (\ensuremath{B_{2g}} ) channel is generated by the exchange interactions along the tetragonal axes (diagonal directions) [Fig.~\ref{fig3}(a)]. 
The observation of strong 2M signals in both channels therefore provides  evidence for the presence of significant exchange interactions (i.e., $J_3$ and $J_{2a,b}$) in \ce{La3Ni2O7}\ along the Ni--O bonding and its diagonal directions.

\begin{table}[b]		
\caption{\label{tab1}%
Parameters used for the 2M calculation. Magnetic moments ($m_{1/2}$) and multiplication factors ($\mathcal{A}_{1/2}$)  are given in  units of \ensuremath{\mu_\mathrm{B}} and $\mathrm{sr}^{-1}\,\mathrm{meV}^{-2}$, respectively.
The exchange interactions and the Lorentzian broadening factor are given in the unit of meV. $S_1 = m_1/g\mu_\mathrm{B} = 0.33$ with a $g$-factor of 2.}
\begin{tabular*}{\columnwidth}{@{\extracolsep{\fill}}ccccc}
\hline\hline	$m_1$	&	$m_2$	&	$\mathcal{A}_1$	&	$\mathcal{A}_2$	&	$\Gamma_0$		\\	\hline
	0.66		&	0.033		&	4.03		&	2.01		&	1.00	\\ \hline\hline			 
\end{tabular*}
\begin{tabular*}{\columnwidth}{@{\extracolsep{\fill}}cccc}
		$J_cS_1$	&	$J_{2a}S_1$	&	$J_{2b}S_1$	&	$J_3S_1$	\\	\hline
		52.8$\pm$6.94		&	1.98$\pm$0.26		&	0.99$\pm$0.19		&	4.62$\pm$0.96		\\ 
\hline\hline\end{tabular*}
\end{table}

To verify whether the observed Raman signals originate from the 2M excitations, we calculate the 2M Raman response using parameters listed in Table~\ref{tab1}, incorporating the high ($m_1$) and low ($m_2$) spin moments consistent with the recent $\mu$SR and NPD studies \cite{plokhikh25arxiv, khasanov25np}. 
We first compute the non-interacting 2M density of states (DOS) spectra, $\Im \overline\chi_0$, based on the $\bf{k}$-dependent magnon energies and the corresponding scattering matrix elements derived from Eq.~(\ref{eq1}) [Fig.~\ref{fig3}(b)]. The final interacting response ($\Im \chi_\textrm{int}$)  is then obtained by applying a random phase approximation (RPA)-type renormalization that accounts for magnon-magnon interactions [Figs.~\ref{fig1}(e),(f)] \cite{fleury68prl, fleury70prl, lockwood87prb, elliott68prl, elliott69jpc}:
\begin{equation}	\label{eq2}	
\Im \chi_\textrm{int} \propto \Im \frac{\overline\chi_0}{1+J_\alpha\cdot\overline\chi_0}.
\end{equation}
Here, $J_3$ and $J_{2a}$ are used for the interaction vertex $J_\alpha$ in the \ensuremath{B_{1g}}\ and \ensuremath{B_{2g}}\ channels, respectively. The full theoretical procedures are outlined in Appendix~\ref{app:e}.

The calculated bare and renormalized spectra are presented in Fig.~\ref{fig3}(c) and compared with the experimental data at 4.5~K. 
Notably, the theoretical results successfully reproduce the key experimental features, i.e., the peak energies, the overall lineshapes, and even the relative intensities of the \ensuremath{B_{1g}}\ and \ensuremath{B_{2g}}\ signals. This close agreement validates our assignment of the \ensuremath{B_{1g}}\ and \ensuremath{B_{2g}}\ features to the 2M excitations and, in turn, confirms the presence of significant in-plane exchange interactions along both the bonding and diagonal directions.

These results constitute the first spectroscopic evidence for two distinct intralayer exchange interactions: $J_3$ along the Ni--O bond direction and $J_{2a,b}$ along the diagonal. 
$J_3$ is a superexchange interaction between high-moment ($m_1$) sites mediated by an intervening low-moment ($m_2$) site, while $J_{2a,b}$ represents a direct $\pi$-hopping exchange between nickel \ensuremath{d_{x^2-y^2}}\ orbitals \cite{labollita24prm}. 
Our observations therefore highlight the crucial role of \ensuremath{d_{x^2-y^2}}\ orbitals in the intralayer interaction in forming the SDW ground state, complementing the dominant interlayer magnetic coupling ($J_c$) mediated by \ensuremath{d_{3z^2-r^2}}\ orbitals. 
This significant spin exchange along the diagonal directions has not been directly identified by previous RIXS and INS studies \cite{chen24nc, xie24sb}.
 Importantly, the presence of bond-diagonal spin exchange interactions marks a key distinction between the magnetism of \ce{La3Ni2O7}\ and that of the cuprates.

Despite the system's metallicity,   \ce{La3Ni2O7}\ exhibits  pronounced 2M Raman signals that are well described  by the FL exchange-scattering mechanism.
It is noted that  the metallic behavior does not strictly preclude the 2M scattering.
For example, metallic \ce{La_{2-x}Sr_{x}CuO_4}   exhibit conspicuous 2M Raman signals, even though their intensities decrease with increasing hole doping $x$ \cite{sugai03prb, onose04prb, sugai13jpcm}.
\ce{La3Ni2O7}\ ($\sim10^{-2}-10^{-1}~\Omega\cdot\mathrm{cm}$) is comparable to, or slightly more resistive than, typical metallic \ce{La_{2-x}Sr_{x}CuO_4} ($\sim10^{-3}-10^{-2}~\Omega\cdot\mathrm{cm}$) at room temperature \cite{taniguchi95jpsj, ando04prl}.
Given that the bilayer nickelate is less metallic than the hole-doped cuprates, the nickel spins in \ce{La3Ni2O7}\ are expected to be sufficiently localized to support robust local spin-exchange excitations.

In contrast, the 2M excitations of trilayer \ce{La4Ni3O10}\ ($d^{7.33}$) were barely observable in the low-temperature Raman data \cite{gim25prl}, consistent with the fact that the trilayer nickelate is an order of magnitude more metallic ($\sim10^{-4}-10^{-3}~\Omega\cdot\mathrm{cm}$) \cite{zhang20nc} than both \ce{La3Ni2O7}\ and \ce{La_{2-x}Sr_{x}CuO_4}.
Therefore, the presence of these prominent 2M modes further distinguishes \ce{La3Ni2O7}\ from other layered nickelates such as \ce{La4Ni3O10}, indicating that \ce{La3Ni2O7}\ is positioned  closer to the localized, Mott-insulating limit.

Additionally, our analysis reveals that a competition between two distinct diagonal AF interactions is essential to the magnetism in \ce{La3Ni2O7}. 
 To reproduce the lineshape of the high-energy \ensuremath{B_{2g}}\ 2M peak, the model requires two diagonal AF interactions: $J_{2a}$ between the opposite magnetic sublattices and $J_{2b}$ within the same sublattice [Fig.~\ref{fig3}(a)]. 
 As demonstrated in Appendix~\ref{app:e}, if $J_{2b}$ is absent or ferromagnetic, it fails to reproduce the observed \ensuremath{B_{2g}}\ spectrum. 
 Furthermore, the magnetic structure with $\bf{Q}_\mathrm{SDW}$ becomes unstable if $J_{2a} < J_{2b}$. 
 Consequently, in order to consistently reproduce the relative 2M spectra while preserving the stability of the reported $\bf{Q}_\mathrm{SDW}$, our calculations constrain $J_{2a}$ to be approximately twice as large as $J_{2b}$.

 The presence of competing diagonal AF interactions provides a natural explanation for stabilizing spin-stripe order at low pressures and has profound implications for the high-pressure tetragonal phase where SC emerges. 
 In the low-pressure orthorhombic phase, the lattice distortion lifts the degeneracy of the two diagonal directions ($x' \neq y'$). 
 The resulting imbalance between $J_{2a}$ and $J_{2b}$ allows a uniaxial AF stripe pattern propagating along the diagonal direction determined by the stronger AF interaction path. 
 In the high-pressure tetragonal phase, the orthorhombic distortion vanishes so that $J_{2a}$ and $J_{2b}$ should become equivalent. 
 Given that the on-axis interaction ($J_3$) likely remains dominant, the presence of two degenerate and competing diagonal AF interactions in the tetragonal phase is expected to induce strong in-plane spin frustration, which could result in significant spin fluctuations and potentially destabilize the long-range stripe order.

\section{Collective low-energy magnetic modes} \label{sec:4}

Beyond the prominent high-energy 2M peaks, we identify distinct features emerging below 10~meV in the \ensuremath{B_{1g}}\ and \ensuremath{B_{2g}}\ spectra at $T< 30$~K [insets of Figs.~\ref{fig2}(b),(c)]. 
These features are reproduced by our theoretical model [Fig.~\ref{fig3}(c) inset], corresponding to 2M excitations originating from the low-energy magnon branches below 5~meV in the spin wave dispersion [Fig.~\ref{fig3}(b)]. 
Crucially, these low-energy 2M peaks vanish in the calculations if we assume either an ideal spin-charge stripe ($m_2 = 0$) or a simple double-spin stripe with uniform moments ($m_2 = m_1$), both of which have been recently considered as possible ground state configurations \cite{chen24nc} (see the additional calculation results in Appendix~\ref{app:e}).

Based on the calculations with varying $m_2$, the value of $m_2=0.033\mu_\mathrm{B}$ is determined in order to  reproduce the observed Raman data, as in Fig.~\ref{fig3}(c).
While the value of $m_2=0.05\mu_\mathrm{B}$ reported from the $\mu$SR/NPD study \cite{plokhikh25arxiv} also reasonably produces the spectra close to the Raman measurement results, it is found that $m_2=0.033\mu_\mathrm{B}$ yields a better simulation of the observed lineshapes and the peak energies in the low-energy ranges.
While our Raman data support the presence of spin disproportionation, the precise value of $m_2$ should be more directly obtained from $\mu$SR/NPD measurements, because the dynamic Raman response of local spin-flips can be additionally influenced by higher-order interactions and/or thermal effects.

Therefore, the observation of the 2M excitations below 10~meV constitutes  spectroscopic evidence for small but finite low-spin moments, corroborating the recent suggestions from the NPD and $\mu$SR analyses \cite{plokhikh25arxiv, khasanov25np}. 
By unveiling these symmetry-specific, collective excitations, our results support the conclusion that the double-spin stripe with a nonuniform spin moment distribution ($m_1\approx20m_2\approx0.66\mu_\mathrm B$) is the true spin ground state in \ce{La3Ni2O7}.

The assignment of the low-energy feature to the 2M excitations arising from the low-moment magnetic sublattices aligns with the RIXS and XAS reports indicating mixed $d^8 + d^8\underline{L} + d^7$ valence states ($\underline{L}$ denotes ligand holes) in \ce{La3Ni2O7}\ \cite{chen24nc}. 
Note that the mixed valence state in the perovskite nickelates is known as the origin of mixed spin states (i.e., spin disproportionation) \cite{piamonteze05prb}. 
Based on these converging results, we conclude that our spectroscopic data provide  evidence for spin disproportionation in the magnetic structure of \ce{La3Ni2O7}. 
As a result, the system exhibits two distinct spin moments, with the weaker moment being approximately an order of magnitude smaller than the stronger one.

\section{Incipient lattice instability} \label{sec:5} 

\begin{figure}[b] 		
{\includegraphics[width=.9\columnwidth,draft=false]{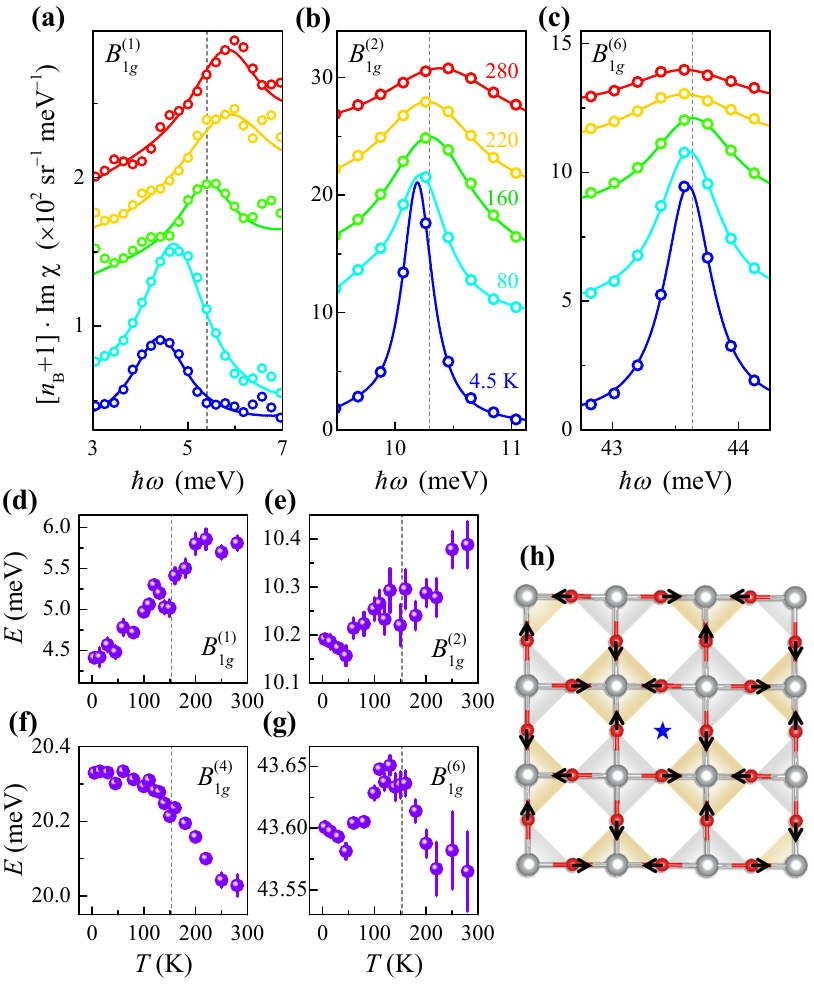}} \vspace{-1em}
\caption{\label{fig4}%
Soft \ensuremath{B_{1g}}\ phonons. 
(a)-(c) Raman spectra of the $B_{1g}^{(1)}$,  $B_{1g}^{(2)}$, and $B_{1g}^{(6)}$ phonon peaks at representative temperatures (plotted with vertical offsets and without Bose correction). Vertical dashed lines mark the phonon energy at 160~K. 
(d)-(g) Temperature-dependent energies of $B_{1g}^{(1)}$,  $B_{1g}^{(2)}$,  $B_{1g}^{(4)}$, and  $B_{1g}^{(6)}$ phonons. Vertical dashed lines indicate \ensuremath{T_\mathrm{SDW}}. Data for other phonons are shown in Fig.~\ref{fig-8}. (h) Atomic displacements for the octahedral breathing mode (arrows), which are inverted under a 90\textdegree\ rotation (indicated by the blue star) representing the \ensuremath{B_{1g}}\ symmetry.}
\end{figure}

As summarized in Figs.~\ref{fig4}(a)-(g), the phonon modes in \ce{La3Ni2O7}\ exhibit distinct temperature dependence. 
While most optical phonons display conventional anharmonic hardening upon cooling [Fig.~\ref{fig4}(f); Appendix~\ref{app:b}], several modes exhibit anomalies clearly linked to the magnetic transition. For instance, the $B_{1g}^{(6)}$ mode initially hardens at high temperatures before softening below \ensuremath{T_\mathrm{SDW}}\ [Fig.~\ref{fig4}(g)], and certain \ensuremath{A_{1g}}\ phonons (e.g., $A_{1g}^{(11)}$ and $A_{1g}^{(16)}$) display kinks in their temperature dependence at \ensuremath{T_\mathrm{SDW}}\ (Appendix~\ref{app:b}). This behavior is characteristic of spin-phonon coupling present in \ce{La3Ni2O7}, by which the onset of the long-range magnetic order modifies the lattice dynamics.

Notably, we observe that low-energy $B_{1g}^{(1)}$ and $B_{1g}^{(2)}$ phonons soften continuously upon cooling starting far above \ensuremath{T_\mathrm{SDW}}\ [Figs.~\ref{fig4}(d),(e)]. 
This softening is particularly dramatic for the $B_{1g}^{(1)}$ mode, whose energy decreases smoothly across the SDW transition. The total amount of this energy reduction exceeds 1.5~meV, which is over 25\% of its phonon frequency at room temperature ($\sim$6~meV) and an order of magnitude larger than the shifts in other phonons. The magnitude of this softening, combined with its insensitivity to the magnetic ordering, indicates that this phenomenon is not a simple consequence of spin-phonon coupling, but signals the presence of an underlying lattice softening driven by a separate degree of freedom.

We interpret this pronounced lattice softening as the experimental signature of an incipient lattice instability. Recent theoretical calculations predict that the spin disproportionation in \ce{La3Ni2O7}\ can drive a static checkerboard-type lattice distortion (space group $Pmnm$), characterized by an imaginary mode frequency of the unstable oxygen breathing phonon \cite{zhang25prb}. 
This in-plane oxygen breathing mode corresponds specifically to the \ensuremath{B_{1g}}\ representation [Fig.~\ref{fig4}(h)]. Consequently, the dramatic softening observed in the $B_{1g}^{(1)}$ mode suggests that the dynamic tendency toward the Ni--O bond disproportionation, manifesting as oxygen breathing modulations, persists in \ce{La3Ni2O7}\ from 280 down to 4.5~K. In contrast to the theoretical prediction, however, our experimental results show that the soft $B_{1g}^{(1)}$ mode remains at a finite frequency, suggesting that the Ni--O bond disproportionation  remains dynamic without freezing into a static checkerboard structure. This interpretation is consistent with the reported persistence of the $Amam$ phase at ambient pressure rather than a transition to the lower-symmetry $Pmnm$ phase \cite{zhang25prb}.

Furthermore, this dynamic Ni--O bond disproportionation provides a natural microscopic mechanism for the site-selective magnetic state. By favoring inequivalent, long and short Ni--O bond environments, it modulates the ligand-hole hybridization, thereby resulting in the two different nickel spin moments confirmed by our 2M data. Our work thus offers evidence for a latent, dynamic lattice instability that persists even in the low-temperature regime where the site-selective magnetic order is stabilized. The intrinsic tendency towards the bond disproportionation indicates that the complete description of the nickelate superconductors cannot be based on the magnetism alone, but must explicitly account for the coupling between this proximate electronic instability and the magnetic order as a fundamental ingredient.

\section{Conclusions} \label{sec:6} 

In this work, we investigate the evolution of the coupled electronic, magnetic, and phononic excitations in \ce{La3Ni2O7}\ crystals. Based on symmetry-resolved analysis and model calculations, we establish the presence of two competing, direct spin exchange interactions ($J_{2a}$ and $J_{2b}$) along the diagonal directions, identifying a key distinction between these layered nickelates and the cuprates. In the bilayer nickelates, three in-plane exchange interactions \emph{inherently} compete with one another, leading to diagonal spin stripe propagation in the low-pressure orthorhombic phase and likely driving magnetic frustration in the superconducting tetragonal phase. Our findings demonstrate that, while nickel \ensuremath{d_{3z^2-r^2}}\ orbitals enforce the {static AF bilayer configuration}, the intralayer magnetism {at ambient pressure} is governed by an intricate competition between the in-plane spin interactions mediated by the \ensuremath{d_{x^2-y^2}}\ orbitals.

\begin{figure*}[] 		
{\includegraphics[width=1.6\columnwidth,draft=false]{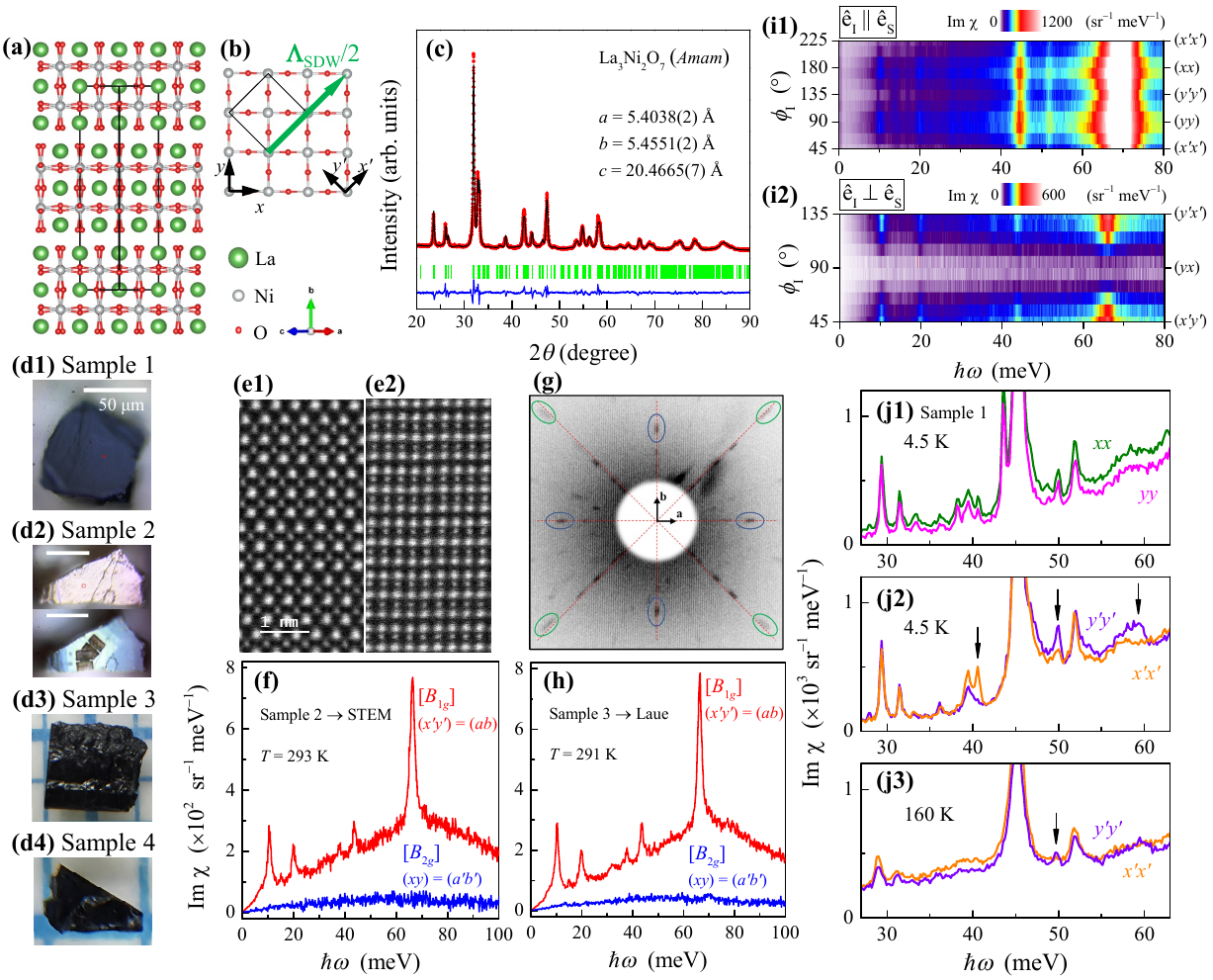}} \vspace{-.8em}
\caption{\label{fig5}%
Crystal structure and angular dependence. 
(a) Crystal structure of \ce{La3Ni2O7}. 
(b) Schematic of a single \ce{NiO2} plane, defining the tetragonal ($x$, $y$) and rotated ($x'$, $y'$) axes for Raman measurements, and the orientation of the halved SDW ordering vector, ${\Lambda}_\mathrm{SDW}/2$ (green arrow). The orthorhombic unit cell is indicated by the black solid lines in (a) and (b). 
{(c) XRD refinement result on ground \ce{La3Ni2O7}\ crystals.
(d1)-(d4) Optical images of the \ce{La3Ni2O7}\ crystals used in this study, denoted as Samples~1 -- 4, respectively.
White scale bars denote 50~$\mu$m, and blue paper grids represent 1~mm intervals.}
(e1),(e2)  HAADF STEM images of the direction-dependent cross-sections of Sample~2 in (d2).
The top and bottom of (d2) are the crystal images before and after fabricating STEM lamellae using FIB along the two cross-sectional directions, respectively.
{(f) Perpendicularly polarized room-temperature Raman spectra of Sample~2, of which the crystal orientation is determined by (e1) and (e2).
(g) Laue measurement result on Sample~3 in (d3). Blue ovals indicate streaks along the orthogonal orthorhombic axes, and green ovals indicate streaks along two Ni--O--Ni bonding directions that are not intersecting at 90\textdegree. Red dotted lines intersecting at the center indicate the 45\textdegree\ intervals.
(h) Perpendicular-polarization room-temperature Raman spectra of Sample~3, with the crystal orientation determined by (g).}
(i1),(i2)  Angular dependence of the room-temperature Raman response for parallel (i1) and perpendicular (i2) linear-polarization configurations as a function of the angle ($\phi_\mathrm{I}$) between the $x$-axis and the incident laser polarization. 
(j1)-(j3) Low-temperature spectra showing subtle evidence of the orthorhombic distortion. While spectra in the ($xx$) and ($yy$) geometries are nearly identical (j1), a few phonon modes  exhibit different intensities in ($x'x'$) versus ($y'y'$) geometries [indicated by the black  arrows in (j2), (j3)], confirming the inequivalence of the two orthorhombic $x'$ and $y'$ axes.
}
\end{figure*}

Furthermore, we identify low-energy magnetic modes corresponding to collective excitations of the small nickel moment ($m_2$), providing strong evidence of spin disproportionation in \ce{La3Ni2O7}. We also find an intriguing anomaly in the phononic sector: an anomalous softening of the \ensuremath{B_{1g}}\ phonons that persists from room temperature down to 4.5~K. As the breathing \ensuremath{B_{1g}}\ mode drives the checkerboard-type lattice modulations, we propose that an incipient bond or charge disproportionation serves as the common underlying driver for both the site-selective spin moments and the incipient lattice instability. This reveals that the SDW ground state is inextricably coupled to lattice softness, a central feature governing the intertwined spin and charge density dynamics in \ce{La3Ni2O7}.

\begin{figure*}[] 		
{\includegraphics[width=1.9\columnwidth,draft=false]{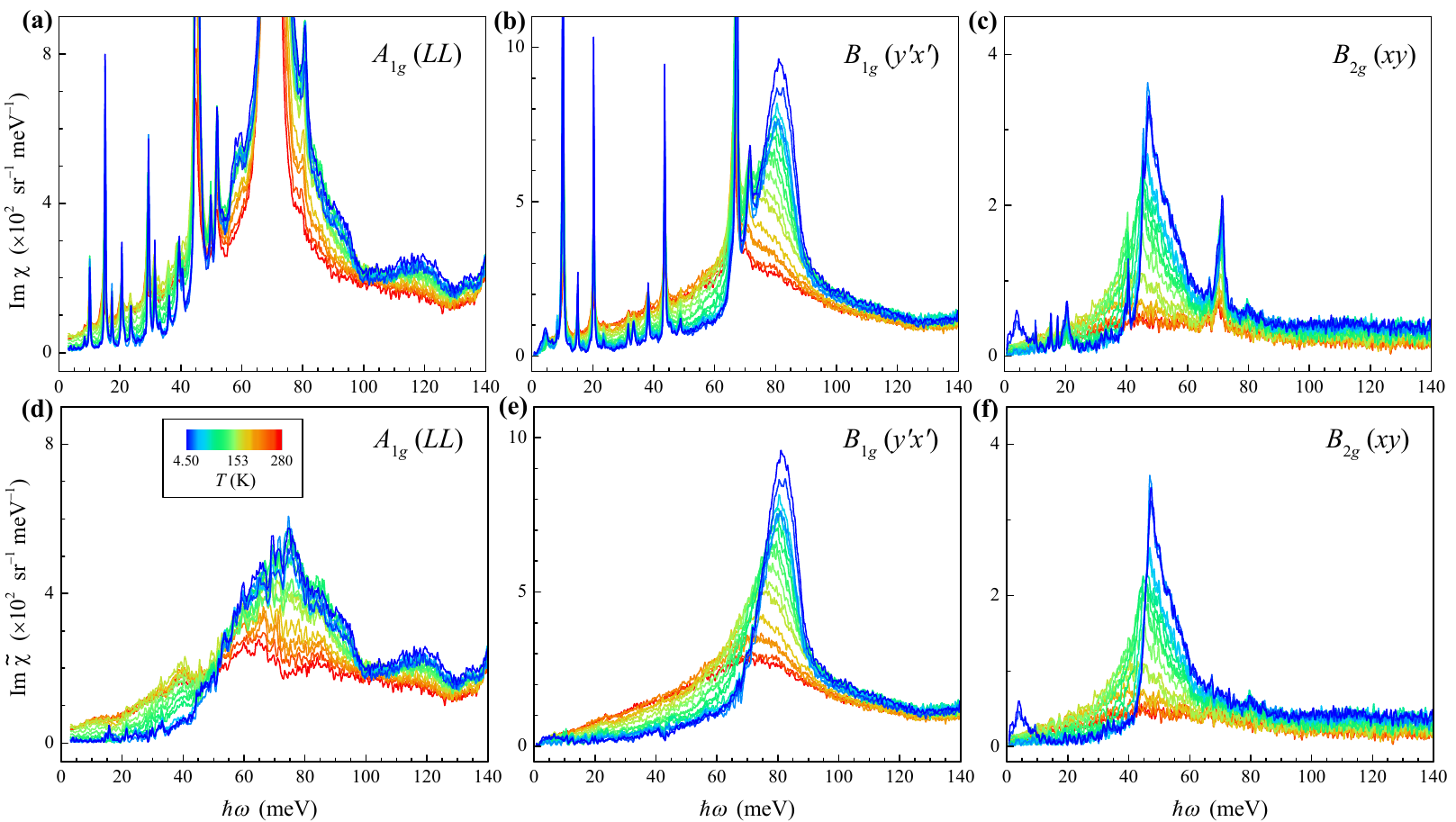}} \vspace{-.8em}
\caption{\label{fig6}%
\ce{La3Ni2O7}\ Raman response before and after phonon subtraction. 
(a)-(c) The raw Raman response ($\Im \chi$) for the \ensuremath{A_{1g}}, \ensuremath{B_{1g}}, and \ensuremath{B_{2g}}\ symmetry channels, respectively. Sharp phonon peaks are superimposed on the broader electronic and magnetic continua. 
(d)-(f) The phonon-subtracted Raman response ($\Im \tilde\chi$) obtained after fitting the phonon peaks to Lorentzian functions and subtracting them from the raw response, which isolates the electronic and magnetic contributions discussed in the main text.
}
\end{figure*}

\begin{acknowledgments}
D.-H.G. and K.H.K. are grateful to Beom Hyun Kim and Duck Young Kim for helpful discussions. This work was supported by the Ministry of Education (Grant No.~2021\allowbreak{}R1\allowbreak{}A6\allowbreak{}C101B418) and by the National Research Foundation (NRF) of Korea grants funded by the Ministry of Science and ICT (MSIT) (Grants No.~RS-2024-00338707, No.~RS-2023-00220471, and No.~2022M3K2A1083855). 
M.W. was supported by the National Natural Science Foundation of China (Grant No.~12425404) and the Guangdong Major Project of Basic Research (Grant No.~2025B0303000004).
D.W. was supported by the ITRC Program through the IITP, and by the Global Research Development Center (GRDC) Cooperative Hub Program through the NRF of Korea, funded by the MSIT (Grants No.~RS-2024-00437191 and No.~RS-2023-00258359).

K.H.K. and M.W. conceived and supervised the project. 
D.H. prepared the samples and performed the XRD measurements. 
H.Z. characterized the Raman spectra of the prepared samples. 
D.-H.G. analyzed the STEM results and performed the Laue measurements. 
D.-H.G. performed the 532-nm Raman measurements, and  
D.-H.G. and D.W. performed the 514.5-nm Raman measurements.
D.-H.G. developed the theoretical models and performed the calculations.
D.-H.G. wrote the original draft. 
D.-H.G., D.W., and K.H.K. reviewed and edited the manuscript with input from all authors.
The authors declare no conflict of interest.
\end{acknowledgments}


\appendix \section{Experimental methods}  \label{app:a}

\ce{La3Ni2O7}\ single crystals were synthesized in an optical floating-zone furnace under a high oxygen pressure of 15~bar. For Raman measurements, fresh crystal surfaces were obtained by mechanical cleavages. 

The macroscopic crystallinity of the grown crystals was confirmed by the powder x-ray diffraction (XRD) measurement result, which was largely refined to the \ce{La3Ni2O7}\ $Amam$ structure as shown in Figs.~\ref{fig5}(a)-(c).
The XRD data contain weak peaks around 24\textdegree, 28\textdegree, and 45\textdegree, which may stem from minor intergrowths of other Ruddlesden--Popper phases (such as \ce{La2NiO4} or \ce{La4Ni3O10}).
The Raman measurements were performed using a micro-focused laser beam on the pure \ce{La3Ni2O7} single crystal pieces, ensuring that the minority phases do not affect the analysis.

The microscopic crystal structure and crystallographic orientation were characterized by scanning transmission electron microscopy (STEM) using a Cs-corrected system (JEM-ARM200F, JEOL) at an acceleration voltage of 200~kV. Cross-sectional lamellae for STEM analysis were prepared using a focused ion beam (FIB) system (Helios Nanolab 650). Both FIB fabrication and STEM measurements were performed at the National Center for Inter-university Research Facilities (NCIRF) at Seoul National University.
The direction-dependent high-angle annular dark-field (HAADF) STEM images are shown in Fig.~\ref{fig5}(e).

To verify the orthorhombic orientation of a  millimeter-scale crystal [Fig.~\ref{fig5}(d3)], the x-ray Laue diffraction measurement was performed using a real-time Laue detector system (MWL110, Multiwire Laboratories) equipped with a tungsten source with an anode voltage of 15~kV.
For the back-reflection Laue experiment, the sample was mounted on a two-stage goniometer installed at a position 111~mm away from the detector.
Fig.~\ref{fig5}(g) presents the Laue measurement result, with blue ovals indicating the two inequivalent, orthogonal $a$ and $b$ axes of the orthorhombic unit cell, and with green ovals indicating the non-orthogonal nearest-neighbor Ni--O--Ni bonding directions in the $ab$-plane.

{Four \ce{La3Ni2O7}\ crystal pieces were used for Raman experiments [Samples~1 -- 4 shown in Fig.~\ref{fig5}(d)].
Sample~1 was used for the main  temperature-dependent measurements.
Samples~2 and 3 were  used for STEM and for Laue measurements, respectively [Figs.~\ref{fig5}(e),(g)], to determine the crystallographic orientation.
Figs.~\ref{fig5}(f),(h) present the perpendicularly polarized Raman spectra of Samples~2 and 3 measured at room temperature, which reveal consistent polarization dependence of the signals. The crystal orientation of Samples~1 and 4 was determined by the polarization dependence of the room-temperature Raman spectra.
To confirm the reproducibility of the thermal evolution observed from Sample~1 discussed in the main text, Samples~3 and 4 were used for obtaining additional temperature-dependent Raman data in Figs.~\ref{fig-17} and \ref{fig-18}, Appendix~\ref{app:f}.}

\begin{figure*}[] 		
{\includegraphics[width=1.9\columnwidth,draft=false]{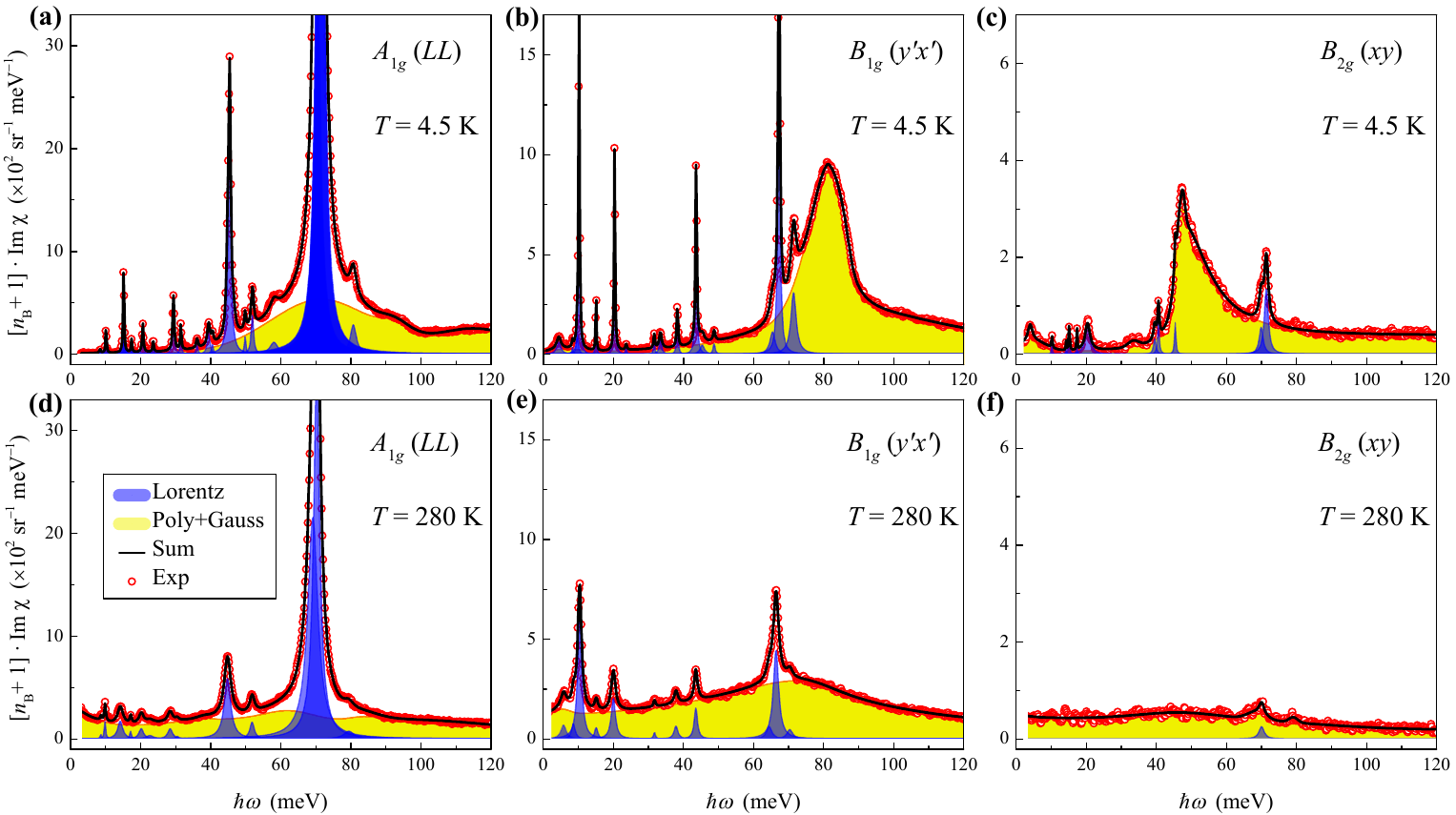}} \vspace{-.8em}
\caption{\label{fig-7}%
{
Representative fits of the phonon peaks in the \ensuremath{A_{1g}}, \ensuremath{B_{1g}}, and \ensuremath{B_{2g}}\ Raman spectra of \ce{La3Ni2O7}\ measured at 4.5 and 280~K. Each phonon peak is fitted within a narrow energy window using a locally defined polynomial baseline (see text for details). Red dots represent the experimental data points, and the blue shaded regions indicate the fitted Lorentzian peak functions for  individual phonon modes. The yellow shaded areas represent the underlying Raman continua, which are modeled using polynomial and Gaussian functions after subtracting all phonon peaks. Black solid lines denote the total sum of the continuum backgrounds and the peak functions.
}
}
\end{figure*}

The temperature-dependent polarized Raman experiments were performed using a commercial micro-Raman system (XperRam200, NanoBase) equipped with a 532-nm laser. The laser power was kept below 0.1~mW to minimize sample heating. An objective lens with a numerical aperture of 0.6 was used to focus the  polarized laser to a spot size $<2~\mu$m on the cleaved \ce{La3Ni2O7}\ crystal surface and to collect the backscattered light. Incident and scattered light polarizations were controlled using half and quarter waveplates and analyzed with linear polarizers. Notch filters were used to reject elastically reflected light. Samples were mounted in a continuous liquid-He cryostat, cooled down to the base temperature of 4.5~K, and warmed up for the temperature-dependent measurements.

The Raman response, $\Im \chi$, was determined from the charge-coupled device (CCD) photocounts by correcting for the Bose–Einstein factor $n_\mathrm{B}(\hslash\omega,T)=(e^{\hslash\omega/k_\mathrm{B}T}-1)^{-1}$  and converting to the absolute unit ($\textrm{meV}^{-1} \textrm{sr}^{-1}$) using the relation:
\begin{equation*}	
\Im \chi(\omega)=\frac{1}{1+n_\mathrm{B}(\hslash\omega,T)} \frac{\pi}{r_0^2}
\frac{\omega_\mathrm{I}}{\omega_\mathrm{S}}
\frac{d^2\sigma}{\hslash d\omega\, d\Omega}.
\end{equation*}
Here, $r_0=\mathrm{e}^2/(4\pi\epsilon_0m_\mathrm{e}c^2)$ is the Thomson electron radius, $\omega_\mathrm{I/S}$ is the incident/scattered light frequency, and $\frac{d^2\sigma}{\hslash d\omega\, d\Omega}$ is the differential Raman scattering cross section calculated from the laser power, the acquisition time, the beam size, the energy shift intervals ($\hslash d\omega$) between CCD pixels, and the solid angle ($\Omega$) of the collecting lens, with instrumental throughput assumed to be constant.

Complementary Raman measurements were performed using a macro-Raman system equipped with a triple-stage spectrometer (TriVista, Princeton Instruments) and a closed-cycle magneto-optical cryostat (SpectromagPT, Oxford Instruments). The Raman data in the out-of-plane magnetic fields $B$ = 0, 7~T at the base temperature (2~K) were collected using a 514.5-nm laser source with a power of 5~mW and a spot diameter of about 50~$\mu$m in the circularly polarized ($RL$) configuration.

\begin{figure*}[] 		
{\includegraphics[width=1.6\columnwidth,draft=false]{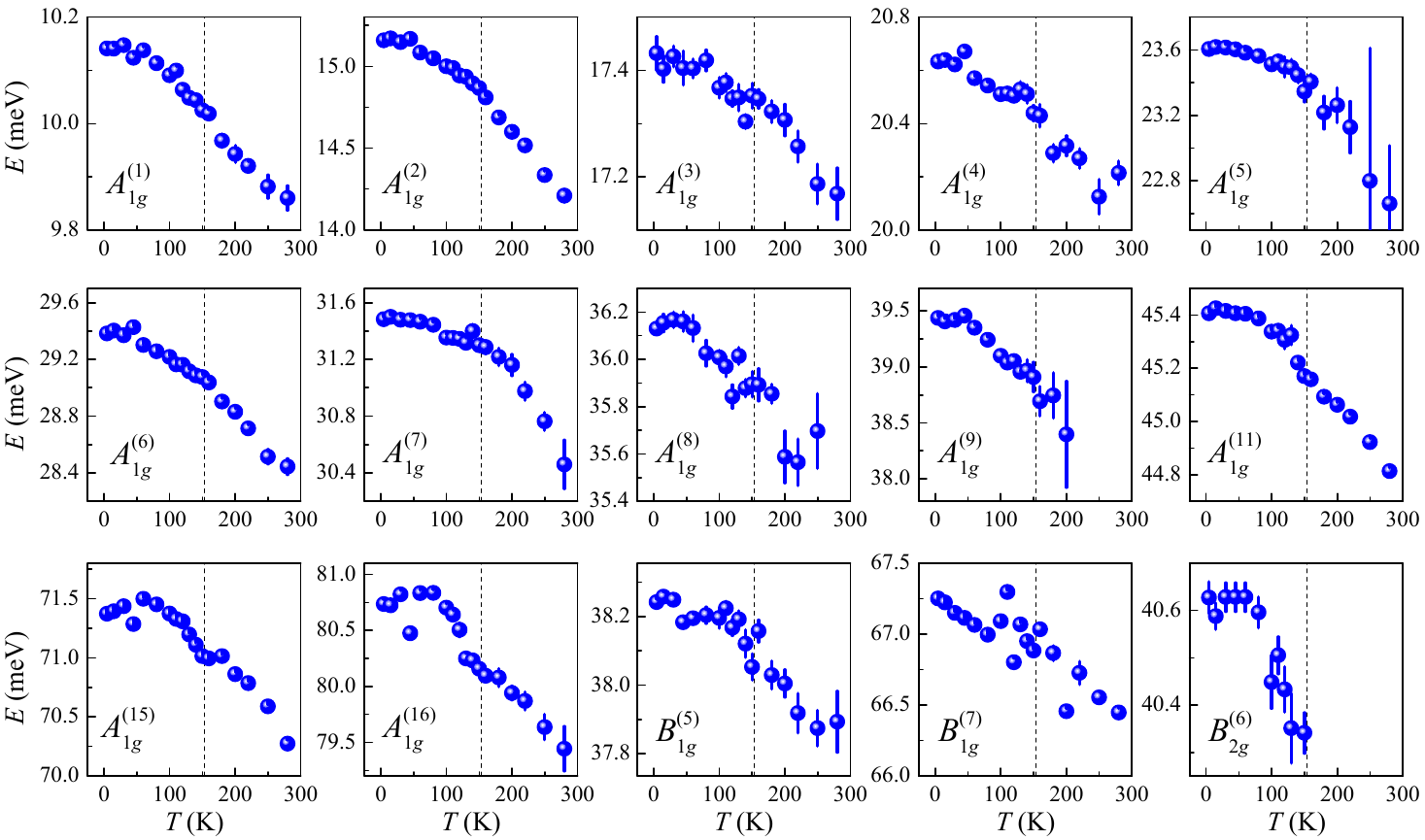}} \vspace{-.8em}
\caption{\label{fig-8}%
Temperature dependence of the energies for representative optical phonons that exhibit typical anharmonic hardening upon cooling, as determined by the Lorentzian fitting analysis. This contrasts with the soft \ensuremath{B_{1g}}\ modes shown in Fig.~\ref{fig4} in the main text. The vertical dashed line indicates \ensuremath{T_\mathrm{SDW}}.
}
\end{figure*}

\section{Analysis and subtraction of phonon peaks} \label{app:b}
To isolate the electronic and magnetic contributions, the phonon peaks were fitted to Lorentzian profiles and subtracted from the spectra, yielding the phonon-subtracted response, $\Im \tilde\chi$. 
The raw and phonon-subtracted Raman responses are compared in Fig.~\ref{fig6}. 
{
To   extract  the underlying electronic and magnetic continuum, the experimental data were simultaneously modeled within a narrow energy window for each peak (or a group of peaks),  using a combination of Lorentzian peak profiles for the sharp phonon modes and a polynomial (linear or quadratic) baseline to capture the broad background.
Representative fits for the highest and the lowest temperatures across each Raman channel are shown in Fig.~\ref{fig-7} (The yellow continuum backgrounds are modeled by the combination of polynomials and Gaussian functions after subtracting all Lorentzian phonon peaks).
}
The temperature dependence of the phonon energies is summarized in Figs.~\ref{fig4}(a)-(d) and Fig.~\ref{fig-8}.

\begin{figure*}[] 		
{\includegraphics[width=1.4\columnwidth,draft=false]{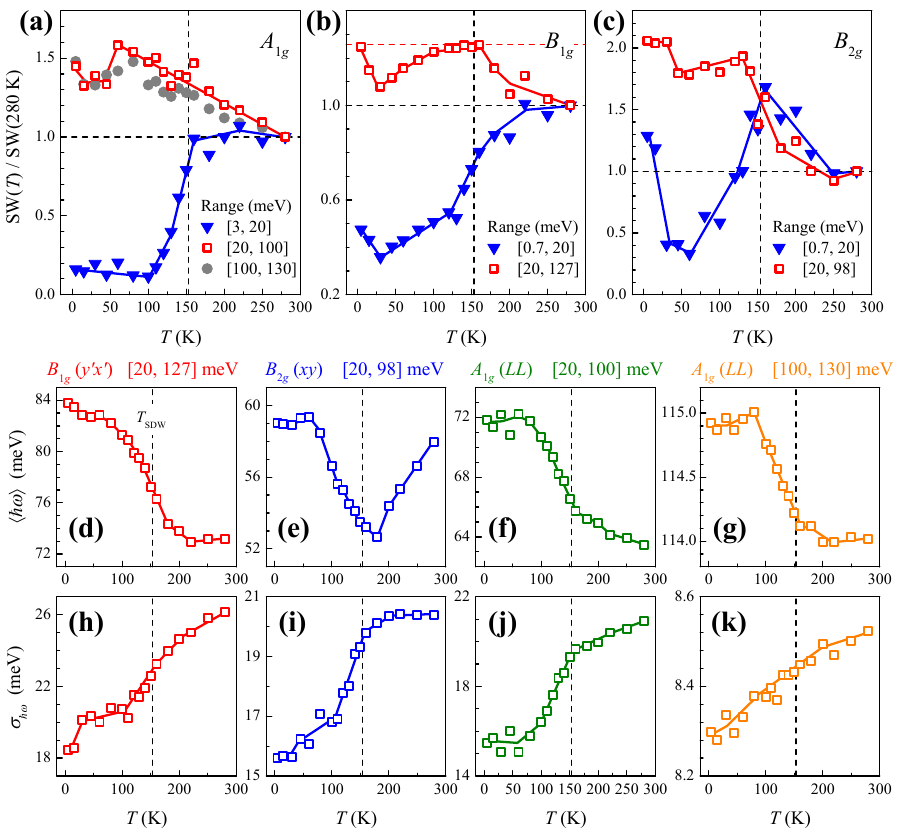}} \vspace{-.8em}
\caption{\label{fig9}%
Temperature dependence of the 2M peak characteristics. (a)-(c) SW evolution for each symmetry channel integrated over the low (blue) and high (red, grey) energy ranges, normalized by the values at 280~K. (d)-(g) Mean energies and (h)-(k) standard deviations (spectral widths) of the 2M spectral features as functions of temperature, calculated over the energy ranges indicated above each panel. The vertical dashed line marks \ensuremath{T_\mathrm{SDW}}. Solid lines are guide for the eye.
}
\end{figure*}

To confirm the validity and robustness of the subtraction procedure, it is verified  that the extracted phonon energies evolve systematically with temperature [Figs.~\ref{fig4}(d)-(g) and \ref{fig-8}].
In addition, it is confirmed that the resulting phonon-subtracted data exhibit highly systematic and smooth spectral-weight, mean-energy, and spectral-width evolutions (Fig.~\ref{fig9}), which validates the reliability of the subtraction method.

\section{Tetragonal approximation} \label{app:c}

In the main text, we analyze our Raman data using the tetragonal $D_{4h}$ point group. We note that the true crystallographic structure of \ce{La3Ni2O7}\ at ambient pressure is orthorhombic ($Amam$, $D_{2h}$) due to slight lattice distortions. However, the experimental evidence demonstrates that the system behaves as effectively tetragonal, making the $D_{4h}$ framework the appropriate and most physically relevant choice for interpreting the data.

Most importantly, the Raman responses in the ($xy$) and ($LL$) scattering geometries are completely different [Figs.~\ref{fig2}(a),(c)]. Under the strict $D_{2h}$ symmetry, both geometries would probe the same $A_g$ representation (Table~\ref{tab2}), and their spectra would be expected to be similar or heavily intermixed. In contrast, under tetragonal $D_{4h}$ symmetry, these geometries select distinct symmetry channels, \ensuremath{B_{2g}}\ and \ensuremath{A_{1g}}, respectively. The pronounced experimental difference between these two channels therefore confirms that any symmetry-lowering effects from the orthorhombic distortion are negligible for the electronic and magnetic excitations central to this study.

\begin{table}[b]		
\caption{\label{tab2}%
Raman-active symmetry channels in each point group.}
\begin{ruledtabular} 
\begin{tabular}{lcc}
	$(\bf{\hat e}_\mathrm{I}\bf{\hat e}_\mathrm{S})$	&	$D_{4h}$	&	$D_{2h}$		\\	\hline
	($xx$), ($yy$)		&	$A_{1g}+B_{1g}$		&	$A_{g}+B_{1g}$			\\
	($x'x'$), ($y'y'$)		&	$A_{1g}+B_{2g}$		&	$A_{g}$			\\
	($xy$), ($yx$)		&	$B_{2g} (+A_{2g})$		&	$A_{g}$			\\
	($x'y'$), ($y'x'$)		&	$B_{1g} (+A_{2g})$		&	$B_{1g}$			\\
	($LL$), ($RR$)		&	$A_{1g} (+A_{2g})$		&	$A_{g}$			\\
	($LR$), ($RL$)		&	$B_{1g}+B_{2g}$		&	$A_{g}+B_{1g}$			\\
\end{tabular}
\end{ruledtabular} 
\end{table}

\begin{figure*}[] 		
{\includegraphics[width=1.5\columnwidth,draft=false]{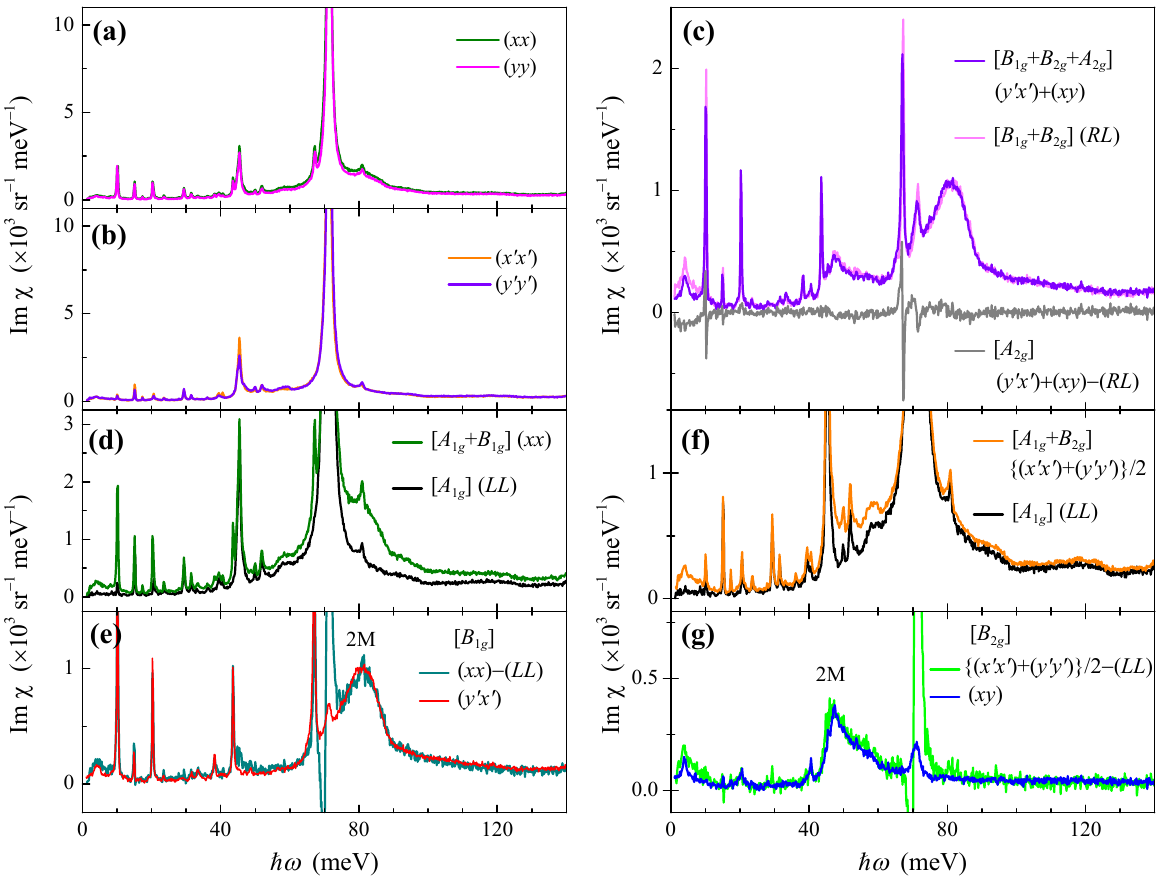}} \vspace{-.8em}
\caption{\label{fig-10}%
Polarization-resolved analysis at 4.5~K. (a),(b) Comparison of the parallel-polarized spectra, between ($xx$) and ($yy$) geometries (a) and between ($x'x'$) and ($y'y'$) geometries (b). 
(c) Isolation of the \ensuremath{A_{2g}}\ signal from the combination of the ($y'x'$), ($xy$), and ($RL$) spectra. The resulting \ensuremath{A_{2g}}\ response (grey) is nearly zero for the entire energy range, except for the residuals from the subtraction of sharp phonon peaks. 
(d),(e) Validation of the \ensuremath{B_{1g}}\ channel assignment. The \ensuremath{B_{1g}}\ response is probed directly from the ($y'x'$) spectrum [red in (e)] and can also be isolated by subtracting the pure \ensuremath{A_{1g}}\ ($LL$) component from the ($xx$) spectrum [black and olive in (d)]. Both methods yield nearly identical spectra for the 2M peak (e), confirming the \ensuremath{B_{1g}}\ assignment. 
(f),(g) Validation of the \ensuremath{B_{2g}}\ channel assignment. Similarly, the \ensuremath{B_{2g}}\ response is measured directly in the ($xy$) spectrum [blue in (g)] and isolated from the $(x'x')+(y'y')$ spectrum [orange in (f)]. Comparison of the both methods (g) demonstrate that the corresponding 2M peak is consistent with the \ensuremath{B_{2g}}\ assignment.
}
\end{figure*}

The effective tetragonal symmetry is further corroborated by the angular dependence of the Raman intensity. At room temperature, the response exhibits a clear four-fold rotational symmetry, with intensity maxima and minima alternating every 45\textdegree\ [Fig.~\ref{fig5}(i)]. 
This is a definitive signature of underlying tetragonal electronic and lattice properties, a behavior also reported in the trilayer nickelate \ce{La4Ni3O10}\ \cite{gim25prl}. While the tetragonal response is dominant, we do observe subtle evidence of the orthorhombic structure in a few low-temperature phonon modes, characterized by different intensities in the ($x'x'$) and ($y'y'$) polarizations [Figs.~\ref{fig5}(j2),(j3)] as $x'$ and $y'$ are two inequivalent orthorhombic axes. Despite these minor modulations of the phonon intensities, the overall Raman response including the two-magnon (2M) features in the \ensuremath{B_{1g}}\ and \ensuremath{B_{2g}}\ channels has nearly four-fold rotational dependence at low temperatures [Figs.~\ref{fig-10}(a),(b)]. This confirms that the tetragonal approximation is robust and provides the insightful basis for the central discoveries in our work: electronic gap, 2M peaks, and soft \ensuremath{B_{1g}}\ phonons.

While bulk crystals may contain twin boundaries, the polarized Raman spectra in this work could be collected from a single orthorhombic domain, owing to the laser beam spot $<2~\mu$m  smaller than  typical twin domains.
The Laue measurement  [Fig.~\ref{fig5}(g)] on a 2-mm crystal [Fig.~\ref{fig5}(d3)] demonstrates that our \ce{La3Ni2O7}\ specimen maintains a single orientation without orthorhombic domain twinning in millimeter length scales. If the crystal contained heavily twinned domains, the Laue measurement would not distinguish the inequivalent orthorhombic axes with different lengths (blue ovals), and eight streaks corresponding to duplicated in-plane Ni--O--Ni bonds would be observed instead of the four green ovals clearly resolved in Fig.~\ref{fig5}(g).
Furthermore, as shown in Fig.~\ref{fig5}(j), the micro-focused polarized Raman measurements  resolve the anisotropic phonon intensities (marked by black arrows), which is a  single-domain characteristic. 
If the laser spot contained mixed twin domains, these peak intensities would be averaged out and appear identical for light polarized along $x'$ and $y'$ directions.

Therefore, we assume in the main analysis that we probe single domains in the Raman experiments.
The nearly four-fold rotational symmetry observed in the overall Raman features is not an artifact of domain averaging, but rather an intrinsic property of the orthorhombic phase. 
Given that a similar quasi-tetragonal Raman response has been reported in the monoclinic trilayer nickelate \ce{La4Ni3O10}\ as well \cite{gim25prl},  the quasi-tetragonal behavior is inferred to be  a fundamental electronic and magnetic property shared by the layered nickelates.

Finally, we also investigated the possibility of an \ensuremath{A_{2g}}\ signal, which would indicate a chiral or time-reversal symmetry-broken state. The \ensuremath{A_{2g}}\ response can be isolated by subtracting a ($RL$) spectrum from the sum of ($xy$) and ($y'x'$) spectra. However, this analysis yielded no discernible evidence for the \ensuremath{A_{2g}}\ signal [Fig.~\ref{fig-10}(c)]. Accordingly, the \ensuremath{A_{2g}}\ channel is not considered in our main analysis.

\section{Assignment of the 2M response} \label{app:d}

\subsection{Exclusion of alternative origins for the \ensuremath{B_{1g}}\ and \ensuremath{B_{2g}}\ peaks}  \label{app:d1} 

In this section, we systematically describe the reasonings to rule out alternative interpretations for the prominent peaks observed in the \ensuremath{B_{1g}}\ and \ensuremath{B_{2g}}\ $\Im \tilde\chi$ spectra.

First, phononic origins can be firmly excluded. A two-phonon scattering process is inconsistent with our data because of symmetry selection rules. The product of two identical irreducible representations (e.g., $B_{1g}\otimes B_{1g}$ or $B_{2g}\otimes B_{2g}$) always contains the fully symmetric \ensuremath{A_{1g}}\ representation, resulting in any two-phonon signal strongest in the \ensuremath{A_{1g}}\ channel. Since the low- and high-energy peaks of the \ensuremath{B_{1g}}\ and \ensuremath{B_{2g}}\ channels are not observed in our \ensuremath{A_{1g}}\ ($LL$) spectra, a two-phonon assignment is untenable. Furthermore, the possibility of a Kohn anomaly related to a CDW instability is also ruled out, as such an anomaly would cause phonon softening. In contrast, the high-energy \ensuremath{B_{1g}}\ and \ensuremath{B_{2g}}\ peaks exhibit hardening below \ensuremath{T_\mathrm{SDW}}.

\begin{figure}[] 		
{\includegraphics[width=.7\columnwidth,draft=false]{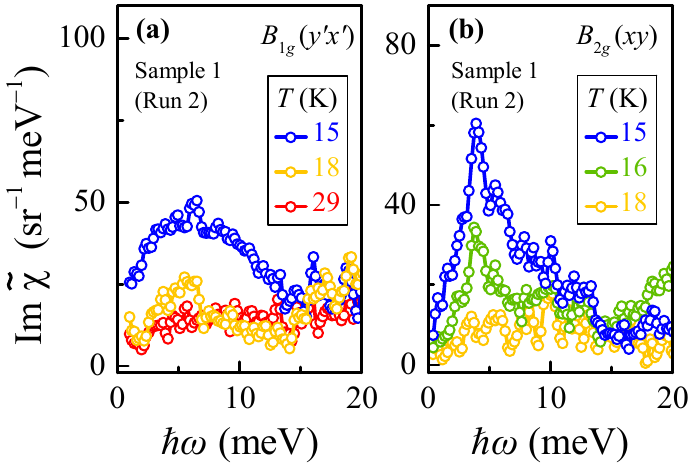}} \vspace{-.8em}
\caption{\label{fig-11}%
{Suppression of the low-energy (a) \ensuremath{B_{1g}}\ and (b) \ensuremath{B_{2g}}\ Raman signals for $T>15$~K.}
}
\end{figure}

The  high-energy 2M peaks (81 and 47~meV) do not originate from  single-phonon excitations of a broken lattice symmetry at low temperatures either, because these modes evolve continuously from room temperature with broad widths $>20$~meV that are  inconsistent with typical widths $<5$~meV of sharp phonon peaks.
Furthermore, the low-energy components below 15~meV in the \ensuremath{B_{1g}}\ and \ensuremath{B_{2g}}\ channels appear only for $T<18$~K (Fig.~\ref{fig-11}).
The lineshapes of the low-energy \ensuremath{B_{1g}}\ and \ensuremath{B_{2g}}\ signals differ from a typical Lorentzian profile of a phonon peak: The \ensuremath{B_{1g}}\ signal exhibits a continuum hump-like structure rather than a sharp peak, whereas the \ensuremath{B_{2g}}\ signal has an asymmetric lineshape, both with   spectral widths extending up to 15~meV.
Besides, in contrast to their rapid suppression above 15~K (Fig.~\ref{fig-11}), no experimental evidence for an ambient-pressure transition that breaks the lattice symmetry has been reported in \ce{La3Ni2O7}\ for this low-temperature range ($T<20$~K) \cite{liu23scpma}.
Given the simultaneous appearance of the non-phonon-like structures in both the \ensuremath{B_{1g}}\ and \ensuremath{B_{2g}}\ channels, as well as their abrupt disappearance just above $T\approx18$~K without an accompanying phase transition, these emergent low-energy Raman components are unlikely to originate from broken-symmetry phonon excitations.

Second, the peaks cannot be attributed to single-magnon excitations due to 
energy considerations and violations of  selection rules. 
The observed peak energies (47 and 81~meV) are inconsistent with the reported single-magnon energy ($\sim$70~meV) at the $\Gamma$-point \cite{chen24nc}.
According to the mechanism established by Elliott, Fleury, and Loudon (EFL), 
a single-magnon excitation is forbidden in parallel  ($xx$) or ($x'x'$) polarization geometries \cite{elliott63pl, fleury68pr}. 
This is inconsistent with the observations of the \ensuremath{B_{1g}}\ and \ensuremath{B_{2g}}\ peaks in such configurations [Figs.~\ref{fig-10}(d)-(g)].
The EFL formalism is applicable if the spin Hamiltonian consists only of isotropic Heisenberg interactions, as in the case of the bilayer antiferromagnetic model   covered in Appendix~\ref{app:e}.

However, the inclusion of other forms of spin interactions in the Hamiltonian may allow $\Gamma$-point magnon excitations to be detected in additional scattering channels.
For example, as demonstrated in \ce{La2CuO4}, anisotropic spin interactions (e.g., $XY$-type) allow single-magnon excitations in the $E_g$ channel (or $B_{3g}$ under the orthorhombic $D_{2h}$ description), while antisymmetric Dzyaloshinskii--Moriya (DM) interactions allow single-magnon excitations in the \ensuremath{B_{1g}}\ channel with the excitation energy corresponding to the DM gap ($\Delta_\mathrm{DM}$) at the $\Gamma$-point \cite{gozar04prl, silvaneto05prb}.

In the \ce{La3Ni2O7}\ spectra, the \ensuremath{B_{1g}}\ and \ensuremath{B_{2g}}\ peaks are observed within the ($xx$), ($x'x'$), ($xy$), and ($y'x'$) scattering geometries [Figs.~\ref{fig-10}(d)-(g)].
This rules out the $XY$-interaction mechanism, which should instead generate signals in the ($xz$) or ($yz$) geometries.
The \ensuremath{B_{2g}}\ magnetic response in  \ce{La3Ni2O7}\ is also inconsistent with the DM-interaction mechanism, as magnons arising from DM interactions should correspond to the \ensuremath{B_{1g}}\ symmetry.
In addition, the \ensuremath{B_{1g}}\ magnetic peak energy detected in \ce{La3Ni2O7}\ (81~meV) is too large to be associated with a typical DM gap scale $\Delta_\mathrm{DM}$.
If the \ensuremath{B_{1g}}\ mode at 81~meV were to correspond to a DM magnon, it would imply that $\Delta_\mathrm{DM}$ exceeds the main Heisenberg $\Gamma$-point magnon energy (70~meV), which is unphysical.
Given the smaller spin moment ($\approx0.66\mu_\mathrm{B}$) of metallic \ce{La3Ni2O7}\ relative to the spin moment ($1\mu_\mathrm{B}$) of insulating \ce{La2CuO4}, as well as the more extended spin periodicity of the ($\pi/2$,$\pi/2$) spin order in \ce{La3Ni2O7}\ compared to the ($\pi$,$\pi$) order in \ce{La2CuO4}, the DM-interaction magnon in \ce{La3Ni2O7}\ is expected to have an energy even smaller than $\Delta_\mathrm{DM}=2.1$~meV of \ce{La2CuO4}.
Therefore, even the \ce{La3Ni2O7}\ magnetic mode near 5~meV, which is detected more prominently in the \ensuremath{B_{2g}}\ response [insets of Figs.~\ref{fig2}(b),(c) and \ref{fig3}(c)], is unlikely to originate from the DM-interaction mechanism, as such a peak should be instead featured dominantly in the \ensuremath{B_{1g}}\ channel with a smaller energy.

Likewise, in $t_{2g}$ systems such as \ce{LaTiO3} \cite{ulrich15prl}, \ce{Sr_{2-x}La_{x}IrO4} \cite{gim16prb, gretarsson17prb}, \ce{Ca2RuO4} \cite{souliou17prl}, and \ce{Li2IrO3} \cite{yang21prb}, single magnon signals can be generated by bond-dependent anisotropic magnetic interactions (e.g. Kitaev interaction) of nearly degenerate $d_{xz}$, $d_{yz}$, and $d_{xy}$ orbitals.
However, because the physics in \ce{La3Ni2O7}\ is determined by the $e_g$ (\ensuremath{d_{x^2-y^2}}\ and \ensuremath{d_{3z^2-r^2}}) orbitals, such bond-directional magnetism arising from spin-orbit entangled $t_{2g}$ states is concluded as irrelevant to the bilayer nickelate.
In conclusion, based on both the EFL and non-EFL mechanisms,  the single-magnon interpretation is inconsistent with the Raman response in \ce{La3Ni2O7}.

Finally, an electronic origin, such as an interband transition across the SDW gap, is inconsistent with the temperature dependence of the spectral weight (SW). The opening of the SDW gap abruptly depletes low-energy SW [Figs.~\ref{fig2}(a),(d)], which would be transferred to higher energies. This requires that the SW of any interband transition related to the SDW gap should increase more steeply below \ensuremath{T_\mathrm{SDW}}\ to compensate for the gapping. However, the SW evolution of the high-energy \ensuremath{B_{1g}}\ and \ensuremath{B_{2g}}\ peaks actually \emph{decreases} below \ensuremath{T_\mathrm{SDW}}\ [red squares in Figs.~\ref{fig9}(b),(c)], in direct opposition to this expectation. The low-energy features below 15~meV are also unlikely to have an electronic origin, because their energy is far below the main SDW gap scale, marked by the isosbestic point near 50~meV. Furthermore, their sharp emergence only below 30~K is incompatible with a Drude response from free carriers, which would be gapped out at the much higher $T_\mathrm{SDW } = 153$~K.

Other ordinary interband transitions present in the paramagnetic state that are unrelated to the SDW ordering can be similarly excluded as follows.
The signal intensity of an ordinary interband transition, even if it is not directly related to the SDW gap itself, is also expected to increase below \ensuremath{T_\mathrm{SDW}}\ due to reduced free charge carriers and suppressed dielectric screening by opening of the  SDW gap.
This is inconsistent with the observed behaviors of the SWs in \ce{La3Ni2O7}, which decrease below \ensuremath{T_\mathrm{SDW}}\  [red squares in Figs.~\ref{fig9}(b),(c)].
Furthermore, the \ensuremath{B_{1g}}\ and \ensuremath{B_{2g}}\ peaks blueshift by 11 and 7~meV, respectively (about 15\% of their absolute peak energies) [Figs.~\ref{fig9}(d),(e)], which is an unexpectedly large energy variance for typical interband energies that have been reported to vary less than $\sim5$~meV \cite{liu24nc}.
More importantly, all \ensuremath{B_{1g}}\ and \ensuremath{B_{2g}}\ peak characteristics change abruptly near \ensuremath{T_\mathrm{SDW}}\ (Fig.~\ref{fig9}), indicating that these excitations are closely related to, and severely affected by, the magnetic transition.

If the \ensuremath{B_{1g}}\ and \ensuremath{B_{2g}}\ peaks resulted from ordinary interband transitions, their peak energies (81, 47~meV) must correspond to gaps between hybridized bands arising from orthorhombic BZ folding given that,  in the tetragonal ($D_{4h}$) phase, all possible direct interband transitions near $E_\mathrm{F}$ have energies greater than 500~meV  \cite{zhang24nc}.
Such zone-folded hybridizations occur primarily for bands crossing with  same point-group symmetries, as isotropic Coulomb interactions ($V\sim1/r$) couple states $\ket{1}$,~$\ket{2}$ with the same symmetry properties, to yield a nonvanishing matrix element with $\mel{1}{V}{2}\ne0$ \cite{dresselhaus}.
Consequently, a low-energy transition across a zone-folded hybridization gap is expected to exhibit the strongest signal intensity in the fully-symmetric \ensuremath{A_{1g}}\ channel (even if it can leak into other channels by the lattice anisotropy), which contradicts the observed  \ensuremath{B_{1g}}\ and \ensuremath{B_{2g}}\ 2M peaks that are completely absent in the \ensuremath{A_{1g}}\ ($LL$) spectra (Fig.~\ref{fig2}).
Therefore, based on the temperature dependence and symmetry constraints, it is concluded that the observed \ensuremath{B_{1g}}\ and \ensuremath{B_{2g}}\ high-energy peaks are  unlikely to be associated with ordinary interband transitions that are not directly related to the SDW transition.

\subsection{Analysis of the detailed temperature dependence}  \label{app:d2} 

In contrast to the scenarios discussed above, the two-magnon (2M) interpretation is fully consistent with our experiments, demonstrated by the excellent agreement between our data and theoretical calculations in the main text. Furthermore, the detailed temperature evolution of the peaks in the \ensuremath{B_{1g}}\ and \ensuremath{B_{2g}}\ channels exhibits the characteristic hallmarks of 2M scattering observed in numerous other antiferromagnets, as we delineate below.

First, the lineshape evolution follows the expected behaviors of the 2M peaks. Above the magnetic ordering temperature, 2M scattering arises from short-range spin fluctuations (paramagnons), typically resulting in broad spectral features. Below the transition, the establishment of long-range order leads to well-defined magnons, causing the 2M peak to both sharpen and harden. This is precisely what we observe in \ce{La3Ni2O7}: Upon cooling below \ensuremath{T_\mathrm{SDW}}, the mean energy of the \ensuremath{B_{1g}}\ and \ensuremath{B_{2g}}\ peaks increases while their spectral width decreases [Figs.~\ref{fig9}(d)-(k)].

Second, the temperature dependence of the integrated SW provides quantitative confirmation. Above \ensuremath{T_\mathrm{SDW}}, the 2M SW depends on the local spin quantum number $S$, decreasing upon cooling for $S \ge 1$ systems but increasing for low-spin $S < 1$ systems \cite{brya74prb}. In \ce{La3Ni2O7}, the SW increases upon cooling above \ensuremath{T_\mathrm{SDW}}\ before peaking near the transition [red squares in Figs.~\ref{fig9}(b),(c)], indicating low-spin $S < 1$ states which is consistent with the previous estimates of nickel spin moments lower than $1\mu_\mathrm{B}$ \cite{plokhikh25arxiv, khasanov25np, zhao25sb}. The subsequent suppression of the SW below \ensuremath{T_\mathrm{SDW}}\ is a consequence of the reduced dynamic spin-spin correlations upon the onset of the magnetic order \cite{brya74prb}, a behavior also seen in other antiferromagnets \cite{brya73aipcp, lockwood87prb}.

Typical 2M excitations often survive in the paramagnetic phase to some extent as broad paramagnon scattering, as in the case for the  high-energy \ensuremath{B_{1g}}\ and \ensuremath{B_{2g}}\ peaks at 81 and 47~meV.
On the other hand, the low-energy \ensuremath{B_{1g}}\ and \ensuremath{B_{2g}}\ components below 15~meV appear only at $T<18$~K [Figs.~\ref{fig2}(b),(c) and \ref{fig-11}].
While the unambiguous interpretation of this behavior requires the development of a rigorous, self-consistent, higher-order formalism for metallic 8-sublattice bilayer antiferromagnets beyond the phenomenological model in Appendix~\ref{app:e}, we note that the low-temperature emergence of the low-energy signals can originate from multiple mechanisms.
For example, since the characteristic energy scale of these signals ($\sim5$~meV) is far smaller than that of the main magnons,  it would necessitate much lower temperatures to overcome thermal broadening and to be resolved as a finite-energy peak.
Because 2M excitations depend strongly on $T$ (as seen from the higher-energy 2M modes),  such weak features in the low-energy range may require $T\ll\ensuremath{T_\mathrm{SDW}}$ to evolve into a well-defined sharp peak.
In addition, the emergence of the low-energy 2M mode should be related to the temperature scale of spin disproportionation in \ce{La3Ni2O7}.
If the smaller spin moments ($m_2$) are stabilized at a temperature far lower than \ensuremath{T_\mathrm{SDW}}\ of the main moments ($m_1$), then the 2M excitations from the weaker spin moment ($m_2$) would accordingly appear  at lower $T$.
Furthermore, when spin disproportionation couples to charge or lattice degrees of freedom (e.g., valence or bond disproportionation),  their complex interplays can modulate the detailed thermal evolution of the 2M signals.
Therefore, local spin-exchange excitations can result in the emergence of the low-energy Raman peaks at $T\ll\ensuremath{T_\mathrm{SDW}}$ in the presence of spin disproportionation.

\subsection{Excitation wavelength and magnetic field dependence}  \label{app:d3} 

\begin{figure}[b] 		
{\includegraphics[width=.7\columnwidth,draft=false]{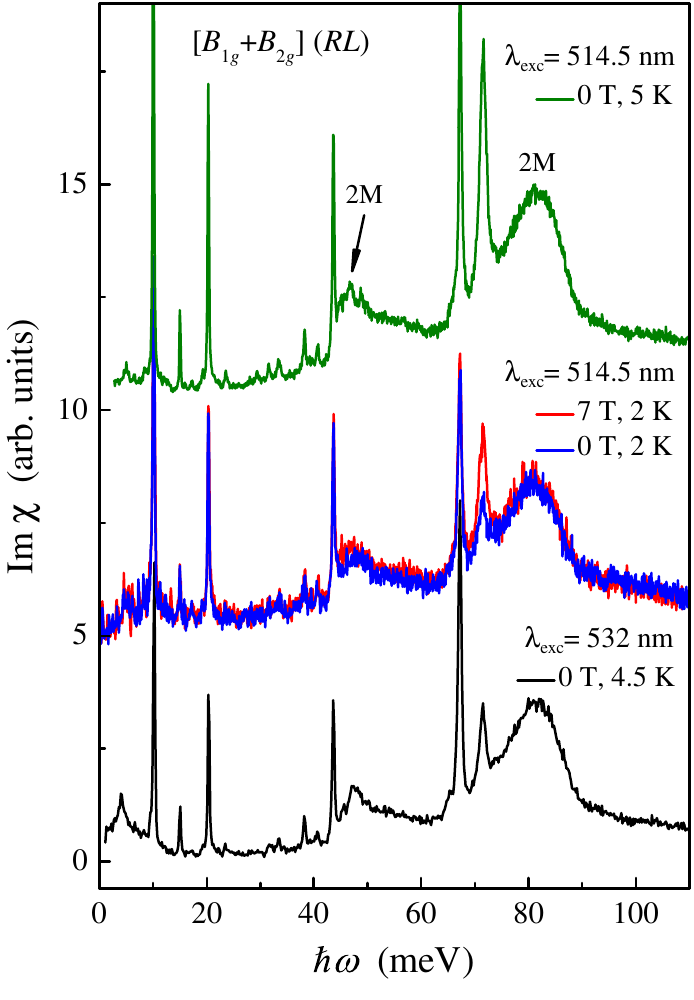}} \vspace{-.8em}
\caption{\label{fig-12}%
Comparison of the low-temperature ($RL$) Raman response measured with two different laser excitation wavelengths (532 and 514.5~nm) and at out-of-plane magnetic fields of $B = 0$ and 7~T (The 514.5-nm data are vertically displaced for clarity). The key spectral features are robust against changes in both wavelength and magnetic field, confirming their origin as 2M Raman scattering.
}
\end{figure}

To further validate our assignment and rule out potential artifacts, we performed complementary measurements using a 514.5-nm laser in a circular ($RL$) polarization geometry, which simultaneously probes the \ensuremath{B_{1g}}\ and \ensuremath{B_{2g}}\ responses, under out-of-plane magnetic fields of $B$ = 0 and 7~T. As shown in Fig.~\ref{fig-12}, the prominent 2M features observed with the 532-nm laser are clearly reproduced in the 514.5-nm data. The persistence of these peaks at a different excitation wavelength provides definitive evidence that they are intrinsic Raman scattering signals and not fluorescent artifacts.

Furthermore, a comparison of the 0 and 7~T data reveals that the 2M features are robust against the applied magnetic field, which is a well-known characteristic of 2M Raman scattering \cite{fleury68pr}. The primary reason is that the 2M process in the Raman scattering involves two opposite spin-flips, whose individual Zeeman energy shifts cancel each other out, leading to no net energy shift. In addition, even in the case of a complete spin-flop phase, the Zeeman energy itself ($\sim$0.4~meV at 7~T) is orders of magnitude smaller than the widths ($> 20$~ meV) of the broad 2M peaks, making any potential shift negligible and hardly observable.

\subsection{Comparison with \ce{La_{2-x}Sr_{x}NiO_4}}  \label{app:d4} 

Given that \ce{La3Ni2O7}\ features  considerable spin-exchange Raman responses, comparison with other layered nickelates exhibiting similar 2M signals, such as \ce{La_{2-x}Sr_{x}NiO_4}, can provide important insights.
Pristine \ce{La2NiO4} ($\mathrm{Ni^{2+}}$, $3d^8$) is an $S=1$ insulator with a simple diagonal ($\pi$,$\pi$) spin-stripe order ($\uparrow$-$\downarrow$-$\uparrow$-$\downarrow$- $\cdots$), and its Raman spectrum features a strong high-energy 2M peak ($\sim200$~meV) in the \ensuremath{B_{1g}}\ channel \cite{sugai90prb, sugai98jpsj}.
Upon hole doping with $x=1/3$ , \ce{La_{5/3}Sr_{1/3}NiO_4} ($d^{7.67}$) develops a mixed spin-charge stripe order ($\uparrow$-$\downarrow$-0-$\uparrow$-$\downarrow$-0-$\cdots$) with  ($2\pi/3$,$2\pi/3$) periodicity formed by coexisting $\mathrm{Ni^{2+}}$ and $\mathrm{Ni^{3+}}$ valence states.
Due to  spin disproportionation, the \ensuremath{B_{1g}}\ Raman spectra of \ce{La_{5/3}Sr_{1/3}NiO_4} features two distinct 2M signals with different energies \cite{yamamoto98prl, blumberg98prl}.

With further doping ($x=1/2$), the half-hole-doped compound \ce{La_{3/2}Sr_{1/2}NiO_4} ($d^{7.5}$) stabilizes the ($\pi/2$,$\pi/2$) spin-charge stripes ($\uparrow$-0-$\downarrow$-0-$\uparrow$-0-$\downarrow$-0-$\cdots$) and exhibits an additional broad Raman peak in the \ensuremath{B_{2g}}\ channel \cite{yamamoto03prb}.
Such an anomalous \ensuremath{B_{2g}}\ mode was conjectured as a magnetic excitation similar to the 2M peaks in the \ensuremath{B_{1g}}\ channel \cite{yamamoto03prb}, which nevertheless could not be concretely verified back then, perhaps due to the lack of detailed information on the exchange interactions in those hole-doped single-layer nickelates.
In contrast, based on the FL+RPA formulations (Fig.~\ref{fig3}; Appendix~\ref{app:e}),  the \ensuremath{B_{1g}}\ and \ensuremath{B_{2g}}\ Raman responses of \ce{La3Ni2O7}\ in the present study are quantitatively verified as originating from the local spin-exchange excitations.

These earlier Raman investigations, reporting that  \ce{La_{2-x}Sr_{x}NiO_4} exhibits the   signatures of multichannel spin exchange pathways (i.e., \ensuremath{B_{1g}}+\ensuremath{B_{2g}}) and spin disproportionation as the nickel valence smoothly changes from $d^8$ to $d^{7.5}$, align  well with the  observations in the bilayer nickelate ($d^{7.5}$)  outlined in the main text. 
The emergence of nickel disproportionation and the diagonal in-plane spin interactions may be  a shared characteristic among these layered nickelates residing on the hole-doped side of  the neutral Ni-$3d^8$ valence. 
While the spin orders in  \ce{La3Ni2O7}\ and in \ce{La_{3/2}Sr_{1/2}NiO_4} are qualitatively analogous to each other, these two systems have a  notable difference.
For example, the \ensuremath{d_{3z^2-r^2}}\ bands in \ce{La_{2-x}Sr_{x}NiO_4} are gapped away from $E_\mathrm{F}$ due to a sizable intraorbital Coulomb interaction $U$ \cite{goodenough82mrb}, whereas the energy separation of the \ensuremath{d_{3z^2-r^2}}\ bands in \ce{La3Ni2O7}\ originates from strong interlayer hybridizations.
The large value of $J_c$ resulting from  such interlayer bonding in the bilayer 
increases the overall energy of \ce{La3Ni2O7}\ magnons.
Consequently, the 2M Raman peak energies are greater in \ce{La3Ni2O7}\ (81 and 47~meV) than in \ce{La_{3/2}Sr_{1/2}NiO_4} (52 and 21~meV) \cite{yamamoto03prb}  despite the identical $d^{7.5}$ average valence of the both nickelates.

\subsection{Discussion of the high-energy \ensuremath{A_{1g}}\ features}  \label{app:d5} 

\begin{figure}[b] 		
{\includegraphics[width=.8\columnwidth,draft=false]{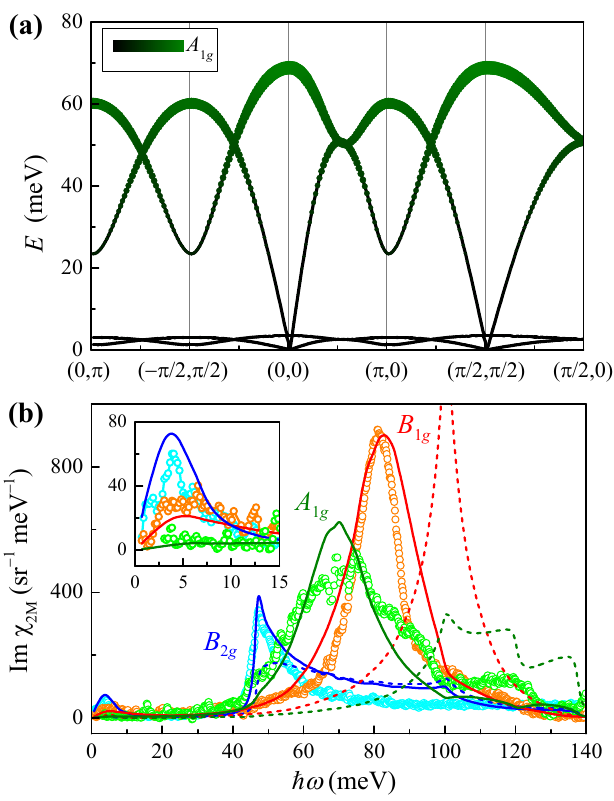}} \vspace{-.8em}
\caption{\label{fig-13}%
Calculation of the 2M peak in the \ensuremath{A_{1g}}\ channel. (a) The \ensuremath{A_{1g}}\ scattering matrix element strength superimposed on the calculated magnon dispersion. (b) Calculated bare (dashed) and interacting (solid lines) 2M responses in the three symmetry channels. To obtain the \ensuremath{A_{1g}}\ response, a phenomenological magnon-magnon interaction vertex of $J_\alpha = 2J_3$ was assumed. The \ensuremath{A_{1g}}, \ensuremath{B_{1g}}, and \ensuremath{B_{2g}}\ data at 4.5~K (green, orange, cyan circles) are compared with the theoretical spectra. High-energy quadratic backgrounds are removed from the \ensuremath{A_{1g}}\ and \ensuremath{B_{1g}}\ data.
}
\end{figure}

Lastly, we address the origin of the broad, high-energy humps observed in the \ensuremath{A_{1g}}\ channel. Their smooth SW evolution across \ensuremath{T_\mathrm{SDW}}\ increasing upon cooling [red and grey in Fig.~\ref{fig9}(a)] might suggest a non-magnetic origin, such as interband electron transitions or multiphonon scattering. However, we find that these features can also be explained by a 2M origin, which become allowed in bilayer systems. Specifically, in a bilayer antiferromagnet, a 2M signal in the \ensuremath{A_{1g}}\ channel can be generated when the interlayer exchange interaction $J_c$ is present, as it breaks the commutation relation between the Heisenberg Hamiltonian ($\mathcal{H}_\mathrm{2LAF}$) and the \ensuremath{A_{1g}}\ ($LL$) Fleury--Loudon scattering operator ($\mathcal{H}_\mathrm{FL}$) \cite{morr96prb}. Indeed, the temperature dependence of the mean energies and spectral widths of the \ensuremath{A_{1g}}\ response resembles that of the \ensuremath{B_{1g}}\ and \ensuremath{B_{2g}}\ 2M peaks [Figs.~\ref{fig9}(d)-(k)].

While the precise form of the magnon-magnon interaction vertex ($J_\alpha$) for the \ensuremath{A_{1g}}\ ($LL$) scattering channel is not obviously known, we performed the model calculation (Fig.~\ref{fig-13}) using a phenomenological assumption of $J_\alpha = 2J_3$. Consequently, the calculation reproduces the key aspects of the high-energy \ensuremath{A_{1g}}\ response, including its spectral distribution and relative intensity [Fig.~\ref{fig-13}(b)]. This interpretation is further supported by observations in the bilayer iridate \ce{Sr3Ir2O7}, where \ensuremath{A_{1g}}\ 2M peaks with a similar temperature dependence have also been reported \cite{li20prl}. We note that a simpler assumption of $J_\alpha = J_c$ fails in this case, as the large value of $J_c$ completely suppresses the renormalized response. 
It can be due to the in-plane light polarizations used in our Raman experiments, not because $J_c$ lacks physical relevance.
The interacting Raman calculations should be viewed as an effective in-plane parameterization, not a fully microscopic treatment controlled by the full exchange hierarchy.
While our results may indicate a 2M origin of the \ensuremath{A_{1g}}\ features, future theoretical works with a more rigorous treatment on the higher-order effects are required for the complete understanding of the high-energy \ensuremath{A_{1g}}\ response.

\section{Theoretical calculation of the 2M response} \label{app:e}

\begin{figure}[] 		
{\includegraphics[width=.6\columnwidth,draft=false]{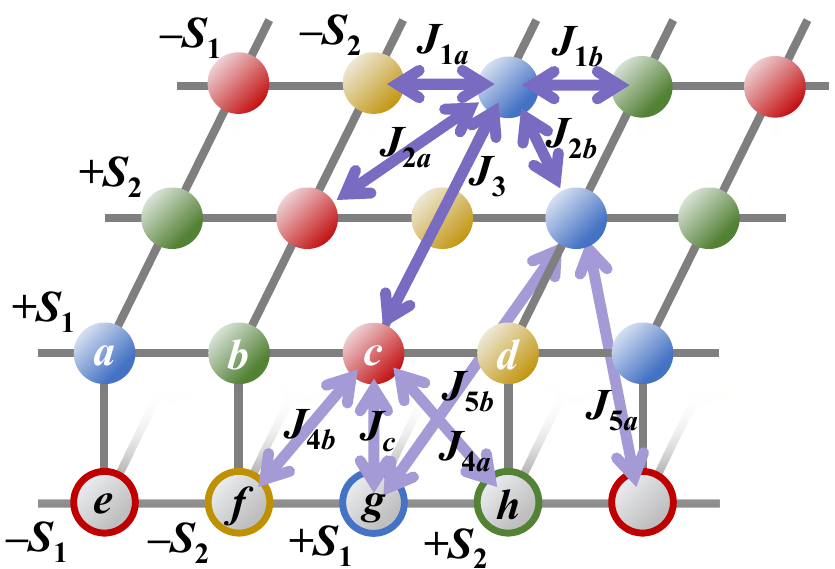}} \vspace{-.8em}
\caption{\label{fig-14}%
Schematic of the magnetic sublattice structure used in the theoretical model, defining the eight distinct nickel sublattices $\{a, b, …, h\}$. The structure consists of high-moment ($\pm S_1$, blue and red) and low-moment ($\pm S_2$, green and orange) sites. The figure illustrates the intralayer ($J_{1a,b}$, $J_{2a,b}$, $J_3$) and interlayer ($J_{4a,b}$, $J_{5a,b}$, $J_c$) exchange interactions included in the Heisenberg Hamiltonian. The subscripts `$a$' and `$b$' distinguish interactions ($J_{1a,b}$ , $J_{2a,b}$, $J_{4a,b}$, $J_{5a,b}$) between sites with antiparallel ($a$) and parallel ($b$) spin alignments, respectively. The interactions not included in Table~\ref{tab1} ($J_{1a,b}$, $J_{4a,b}$, $J_{5a,b}$) are set to zero in the numerical calculation of Fig.~\ref{fig3}.
}
\end{figure}

The magnon dispersion was calculated starting from the bilayer antiferromagnetic (2LAF) Heisenberg Hamiltonian 
\begin{equation}\begin{split}	\label{eq3}	
\mathcal{H}_\mathrm{2LAF}&=\sum_{\langle i,j\rangle}J_{ij}\mathbf{S}_i \cdot\mathbf{S}_j\\ &=\sum_{\langle i,j\rangle}J_{ij}\left[
S_i^z S_j^z + \frac{S_i^+ S_j^-+S_i^- S_j^+}{2}
\right],
\end{split}\end{equation}
where $S_i^z$ is the component along the easy axis $\perp\mathbf{Q}_\mathrm{SDW}$ and $S_i^\pm=S_i^x\pm iS_i^y$. The model incorporates the eight distinct magnetic sublattices $\{a, b, …, h\}$, with the top $\{a, b, c, d\}$ and bottom $\{e, f, g, h\}$ layers coupled antiferromagnetically: $S_1 = S_a = -S_c = -S_e = S_g = m_1/g\mu_\mathrm B$ and $S_2 = S_b = -S_d = -S_f = S_h = m_2/g\mu_\mathrm B$ (Fig.~\ref{fig-14}).

\begin{figure*}[] 		
{\includegraphics[width=1.8\columnwidth,draft=false]{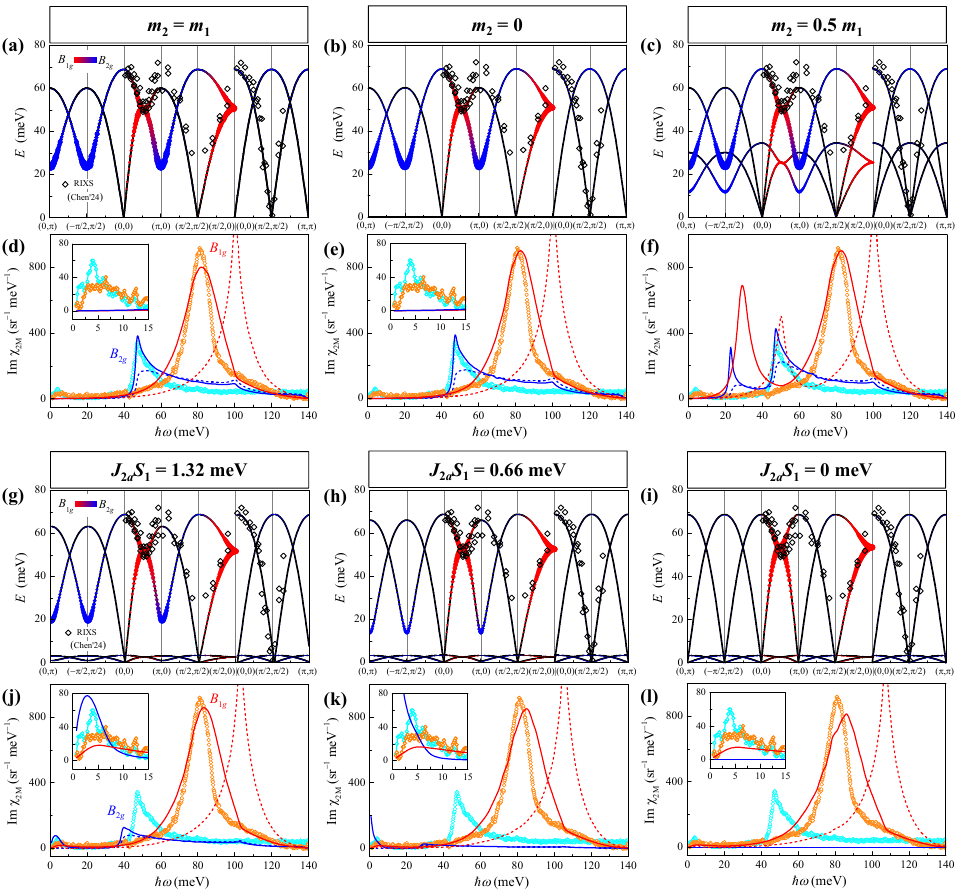}} \vspace{-.8em}
\caption{\label{fig-15}%
Dependence of the 2M response on $m_2$ and $J_{2a}$. 
(a)-(f) Dependence of the 2M response on the low-moment $m_2$. 
For these calculations, all exchange interactions are fixed to the values in Table~\ref{tab1}. 
The magnon band dispersion, the scattering matrix element strengths [(a)-(c)], and the corresponding \ensuremath{B_{1g}}\ and \ensuremath{B_{2g}}\ 2M Raman responses [(d)-(f)] are computed for different values of $m_2$. 
If $m_2$ is zero (b) or made equal to $m_1$ (a), the low-energy features below 15~meV disappear from the calculated spectra [(d),(e) insets]. If $m_2$ is made comparable to $m_1$ (c), additional 2M features appear at a much higher energy range (f), inconsistent with the experimental data. 
(g)-(l) Calculation of the 2M Raman response when the diagonal exchange $J_{2a}S_1$ is set to values lower than 1.98~meV used in the main text (Table~\ref{tab1}). In these calculations, $J_{2b}$ is fixed at half the value of $J_{2a}$, and $J_3S_1$ is adjusted to reproduce the magnon energies reported in the RIXS study \cite{chen24nc}: 4.95~meV (g), 5.28~meV (h), 5.61~meV (i). The resulting spectra show that with decreasing $J_{2a}$, both the intensity and the energy of the \ensuremath{B_{2g}}\ 2M peak decrease and eventually vanish [(j)-(l)]. The colors, symbols, and line styles are identical to those in Figs.~\ref{fig3}(b),(c).
}
\end{figure*}
\begin{figure*}[] 		
{\includegraphics[width=\textwidth,draft=false]{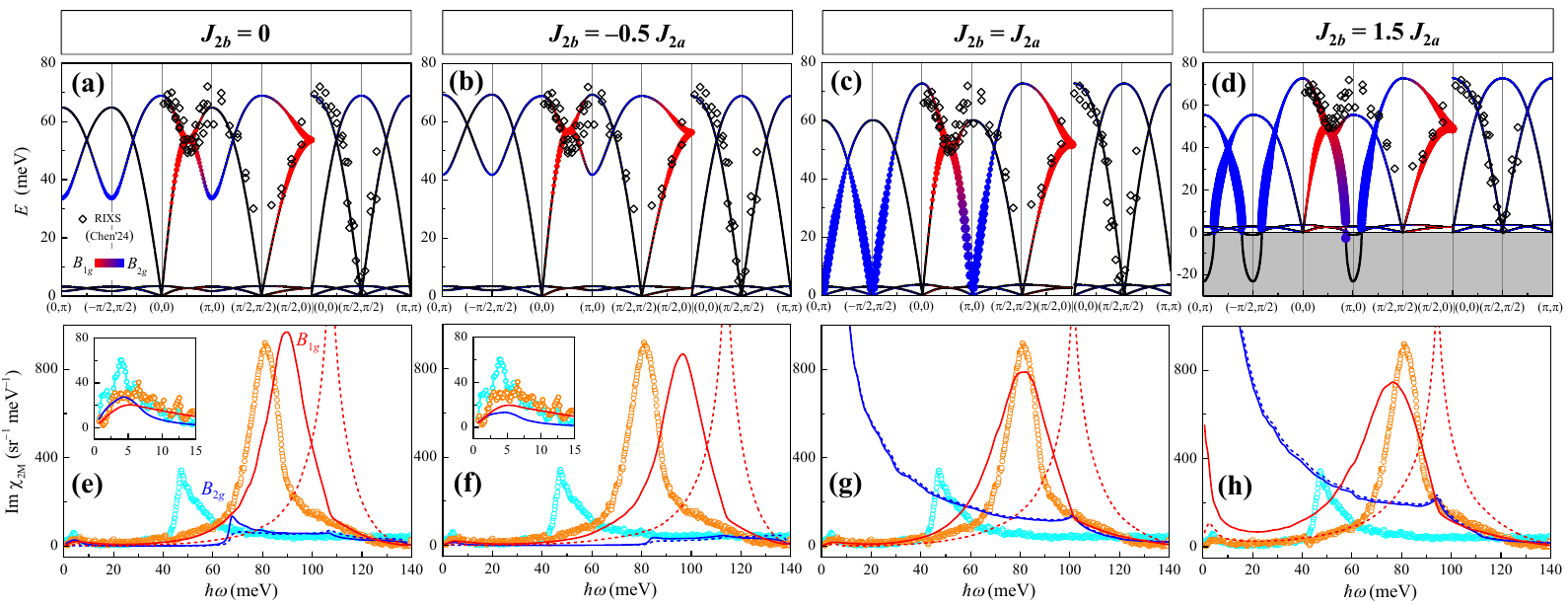}} \vspace{-1.5em}
\caption{\label{fig-16}%
Dependence of the 2M response on $J_{2b}$, the diagonal exchange within each magnetic sublattice. 
(a)-(d) Magnon band dispersion (solid lines) and corresponding scattering matrix element strengths (color intensity) for different values of $J_{2b}$. (e)-(h) The corresponding calculated \ensuremath{B_{1g}}\ and \ensuremath{B_{2g}}\ 2M Raman spectra, compared with the experimental data at 4.5~K. In these calculations, we use exchange and spin parameters given in Table~\ref{tab1}, while we slightly change $J_3S_1$ from 4.62 to 5.28~meV in (c) and (d) to better reproduce the reported magnon energies \cite{chen24nc}. 
For zero and ferromagnetic $J_{2b}$ [(a),(e) \& (b),(g)], the \ensuremath{B_{2g}}\ mode is suppressed and shifted to higher energy, respectively. For larger antiferromagnetic $J_{2b}$ [(c),(g) \& (d),(h)], the \ensuremath{B_{2g}}\ mode gains intensity and moves to lower energy; for example, the energy becomes zero when $J_{2b} = J_{2a}$ [(c),(g)]. If $J_{2b}$ exceeds $J_{2a}$, [(d),(h)], the magnetic structure becomes unstable, resulting in imaginary magnon energies.
}
\end{figure*}

The Hamiltonian was linearized using the leading-order Holstein--Primakoff transformations,
\begin{align*}	       
j\in a&: S_j^+ \simeq \sqrt{2S_1}a_j,\ S_j^- \simeq \sqrt{2S_1}a_j^\dag,\  S_j^z = S_1-a_j^\dag a_j; \\ 
j\in b&: S_j^+ \simeq \sqrt{2S_2}b_j,\ S_j^- \simeq \sqrt{2S_2}b_j^\dag,\  S_j^z = S_2-b_j^\dag b_j; \\ 
j\in c&: S_j^+ \simeq \sqrt{2S_1}c_j^\dag,\ S_j^- \simeq \sqrt{2S_1}c_j,\  S_j^z = -S_1+c_j^\dag c_j; \\ 
j\in d&: S_j^+ \simeq \sqrt{2S_2}d_j^\dag,\ S_j^- \simeq \sqrt{2S_2}d_j,\  S_j^z = -S_2+d_j^\dag d_j; \\* 
\vdots\ & \\ 
j\in h&: S_j^+ \simeq \sqrt{2S_2}h_j,\ S_j^- \simeq \sqrt{2S_2}h_j^\dag,\  S_j^z = S_2-h_j^\dag h_j. 
\end{align*}
Applying the Fourier transformations (e.g., $a_j=\frac{1}{\sqrt{N}}\sum_\mathbf{k}e^{-i\mathbf{k}\cdot\mathbf{r}_j}a_\mathbf{k}$ and $a_j^\dag=\frac{1}{\sqrt{N}}\sum_\mathbf{k}e^{i\mathbf{k}\cdot\mathbf{r}_j}a_\mathbf{k}^\dag$), the Hamiltonian in Eq.~(\ref{eq3}) can be written in matrix form 
$\mathcal{H}_\mathrm{2LAF}=\sum_\mathbf{k}\Phi_\mathbf{k}^\dag K_\mathbf{k} \Phi_\mathbf{k}$, 
where $\Phi_\mathbf{k}=\left(a_\mathbf{k}, b_\mathbf{k}, c_\mathbf{k}^\dag, d_\mathbf{k}^\dag, e_\mathbf{k}^\dag, f_\mathbf{k}^\dag, g_\mathbf{k}, h_\mathbf{k} \right)^T$
is the vector of bosonic operators. The matrix elements of the $8\times8$ dynamical matrix	
\begin{equation} \label{eq4}  
K_\mathbf{k} = \begin{pmatrix}
K_\mathbf{k}^\mathrm{(I)} & K_\mathbf{k}^\mathrm{(II)}\\
K_\mathbf{k}^\mathrm{(II)} & K_\mathbf{k}^\mathrm{(I)}
\end{pmatrix}
\end{equation}
defined for each momentum vector $\mathbf{k}$ by the $4\times4$ matrices
\begin{gather*}
K_\mathbf{k}^\mathrm{(I)}=\begin{pmatrix}
K_\mathbf{k}^{(aa)} & K_\mathbf{k}^{(ab)} & K_\mathbf{k}^{(ac)} & K_\mathbf{k}^{(ad)} \\
K_\mathbf{-k}^{(ab)} & K_\mathbf{k}^{(bb)} & K_\mathbf{-k}^{(ad)} & K_\mathbf{k}^{(bd)} \\
K_\mathbf{k}^{(ac)} & K_\mathbf{k}^{(ad)} & K_\mathbf{k}^{(aa)} & K_\mathbf{k}^{(ab)}\\
K_\mathbf{-k}^{(ad)} & K_\mathbf{k}^{(bd)} & K_\mathbf{-k}^{(ab)} & K_\mathbf{k}^{(bb)} 
\end{pmatrix}\textrm{ and}\\
K_\mathbf{k}^\mathrm{(II)}=\begin{pmatrix}
K_\mathbf{k}^{(ae)} & K_\mathbf{k}^{(af)} & K_\mathbf{k}^{(ag)} & K_\mathbf{k}^{(ah)} \\
K_\mathbf{-k}^{(af)} & K_\mathbf{k}^{(bf)} & K_\mathbf{-k}^{(ah)} & K_\mathbf{k}^{(bh)} \\
K_\mathbf{k}^{(ag)} & K_\mathbf{k}^{(ah)} & K_\mathbf{k}^{(ae)} & K_\mathbf{k}^{(af)}\\
K_\mathbf{-k}^{(ah)} & K_\mathbf{k}^{(bh)} & K_\mathbf{-k}^{(af)} & K_\mathbf{k}^{(bf)} 
\end{pmatrix}
\end{gather*}
are functions of $J_{ij}$, $S_1$, and $S_2$:
\begin{eqnarray*}
    K_\mathbf k^{(aa)} &=& 2J_{2b} S_1 \cos(k_x - k_y) +  2S_2 (J_{1a} - J_{1b} + J_{4a} - J_{4b}) \\
    & & +\ 2S_1 (J_{2a} - J_{2b} + 2J_3 + J_{5a} - J_{5b}) + J_c S_1, \\
    K_\mathbf k^{(bb)} &=& 2J_{2b} S_2 \cos(k_x - k_y)  + 2S_1 (J_{1a} - J_{1b} + J_{4a} - J_{4b}) \\
     & & +\ 2S_2 (J_{2a} - J_{2b} + 2J_3 + J_{5a} - J_{5b}) + J_c S_2, \\
    K_\mathbf k^{(ab)} &=& J_{1b} \sqrt{S_1 S_2} (e^{-ik_x} + e^{-ik_y}),\\
 K_\mathbf k^{(ad)} &=& J_{1a} \sqrt{S_1 S_2} (e^{ik_x} + e^{ik_y}), \\
    K_\mathbf k^{(ac)} &=& 2S_1 (J_{2a} \cos(k_x + k_y) + J_3 (\cos 2k_x + \cos 2k_y)), \\
    K_\mathbf k^{(bd)} &=& 2S_2 (J_{2a} \cos(k_x + k_y) + J_3 (\cos 2k_x + \cos 2k_y)), \\
    K_\mathbf k^{(ae)} &=& S_1 (2J_{5a} \cos(k_x - k_y) + J_c), \\
     K_\mathbf k^{(bf)} &=& S_2 (2J_{5a} \cos(k_x - k_y) + J_c), \\
    K_\mathbf k^{(af)} &=& J_{4a} \sqrt{S_1 S_2} (e^{-ik_x} + e^{-ik_y}), \\
 K_\mathbf k^{(ah)} &=& J_{4b} \sqrt{S_1 S_2} (e^{ik_x} + e^{ik_y}), \\
    K_\mathbf k^{(ag)} &=& 2J_{5b} S_1 \cos(k_x + k_y), \\
    K_\mathbf k^{(bh)} &=& 2J_{5b} S_2 \cos(k_x + k_y).
\end{eqnarray*}

The Hamiltonian $\mathcal{H}_\mathrm{2LAF}$ was diagonalized by a generalized Bogoliubov transformation, which was obtained by solving the eigenvalue problem for the Bogoliubov–de Gennes (BdG) Hamiltonian, $(I_\mathrm{BdG}K_\mathbf{k})v_i=\lambda_iv_i$, where $I_\mathrm{BdG}=\mathrm{diag}(1,1,-1,-1,-1,-1,1,1)$ is the BdG metric. The resulting eigenvalues $\{\lambda_i\}$ give the magnon eigenenergies, $E_i(\mathbf{k})=|\lambda_i|$, and the eigenvectors $\{v_i\}$ construct the corresponding Bogoliubov matrix $T_\mathbf k=(w_1,\ldots,w_8)$ with $w_i=\sqrt{\frac{sgn(\lambda_i)}{v_i^\dag I_\mathrm{BdG}v_i}}v_i$. 
Such a Bogoliubov transformation $T_\mathbf k$ maps the original bosonic operators to the magnon eigenstates: $\Phi_\mathbf k=T_\mathbf k\Phi'_\mathbf{k}$, where
$\Phi'_\mathbf{k}=\left(
\alpha_{1,\mathbf k}, \alpha_{2,\mathbf k}, \alpha_{3,\mathbf k}, \alpha_{4,\mathbf k}, 
\alpha_{5,-\mathbf k}^\dag, \alpha_{6,-\mathbf k}^\dag, \alpha_{7,-\mathbf k}^\dag, 
\alpha_{8,-\mathbf k}^\dag \right)^T$ is the vector of magnon annihilation and creation operators. The Hamiltonian is diagonalized in the magnon eigenbasis $\Phi'_\mathbf{k}$:
\begin{align*}
\mathcal{H}_\mathrm{2LAF}&=\sum_\mathbf{k}\Phi_\mathbf{k}^\dag K_\mathbf{k} \Phi_\mathbf{k}
=\sum_\mathbf{k}{\Phi'}^\dag_\mathbf{k} K'_\mathbf{k} \Phi'_\mathbf{k}\\ &=
\sum_\mathbf{k} \sum_{i=1}^8 E_i(\mathbf{k})\alpha_{i,\mathbf{k}}^\dag \alpha_{i,\mathbf{k}}
\end{align*}
where $K'_\mathbf{k}=T_\mathbf{k}^\dag K_\mathbf{k} T_\mathbf{k}$ 
is the diagonal matrix of magnon eigen-energies.

After applying the same Holstein--Primakoff and Fourier transformations used for Eq.~(\ref{eq3}), the FL scattering Hamiltonian, Eq.~(\ref{eq1}), can also be written in a quadratic bosonic form $\mathcal{H}_\mathrm{FL}=\sum_\mathbf{k}\Phi_\mathbf{k}^\dag M_\mathbf{k} \Phi_\mathbf{k}$, where the scattering vertex matrix 
\begin{equation} \label{eq5}   
M_\mathbf{k} = \begin{pmatrix}
M_\mathbf{k}^\mathrm{(I)} & M_\mathbf{k}^\mathrm{(II)}\\
M_\mathbf{k}^\mathrm{(II)} & M_\mathbf{k}^\mathrm{(I)}
\end{pmatrix}
\end{equation} depends on the light polarizations (i.e., the Raman scattering channel). For example, for the \ensuremath{B_{1g}}\ ($y'x'$) channel, 
\begin{gather*}
M_\mathbf{k}^\mathrm{(I)}=\begin{pmatrix}
0 & K_\mathbf{k}^{(ab)} & M_\mathbf{k}^{(ac)} & M_\mathbf{k}^{(ad)} \\
M_\mathbf{-k}^{(ab)} & 0 & M_\mathbf{-k}^{(ad)} & M_\mathbf{k}^{(bd)} \\
M_\mathbf{k}^{(ac)} & M_\mathbf{k}^{(ad)} & 0 & M_\mathbf{k}^{(ab)}\\
M_\mathbf{-k}^{(ad)} & M_\mathbf{k}^{(bd)} & M_\mathbf{-k}^{(ab)} & 0 
\end{pmatrix},\\
M_\mathbf{k}^\mathrm{(II)}=\begin{pmatrix}
0& M_\mathbf{k}^{(af)} & 0 & M_\mathbf{k}^{(ah)} \\
M_\mathbf{-k}^{(af)} & 0 & M_\mathbf{-k}^{(ah)} & 0 \\
0 & M_\mathbf{k}^{(ah)} & 0 & M_\mathbf{k}^{(af)}\\
M_\mathbf{-k}^{(ah)} &0 & M_\mathbf{-k}^{(af)} & 0 
\end{pmatrix},
\end{gather*}
\begin{eqnarray*}
  M_\mathbf k^{(ab)} &=& J_{1b} \sqrt{S_1 S_2} (e^{-ik_x} - e^{-ik_y}),\\
 M_\mathbf k^{(ad)} &=& J_{1a} \sqrt{S_1 S_2} (e^{ik_x} - e^{ik_y}), \\
    M_\mathbf k^{(ac)} &=& 2  J_3 S_1(\cos 2k_x - \cos 2k_y), \\
    M_\mathbf k^{(bd)} &=& 2  J_3 S_2(\cos 2k_x - \cos 2k_y), \\
      M_\mathbf k^{(af)} &=& J_{4a} \sqrt{S_1 S_2} (e^{-ik_x} - e^{-ik_y}), \\
 M_\mathbf k^{(ah)} &=& J_{4b} \sqrt{S_1 S_2} (e^{ik_x} - e^{ik_y}).  
\end{eqnarray*}
For the \ensuremath{B_{2g}}\ ($xy$) channel,
\begin{gather*}
M_\mathbf{k}^\mathrm{(I)}=\begin{pmatrix}
M_\mathbf{k}^{(aa)} & 0 & M_\mathbf{k}^{(ac)} & 0 \\
0 & M_\mathbf{k}^{(bb)} & 0 & M_\mathbf{k}^{(bd)} \\
M_\mathbf{k}^{(ac)} & 0 & M_\mathbf{k}^{(aa)} & 0\\
0 & M_\mathbf{k}^{(bd)} & 0 & M_\mathbf{k}^{(bb)} 
\end{pmatrix},\\
M_\mathbf{k}^\mathrm{(II)}=\begin{pmatrix}
M_\mathbf{k}^{(ae)} & 0 & M_\mathbf{k}^{(ag)} & 0 \\
0 & M_\mathbf{k}^{(bf)} & 0 & M_\mathbf{k}^{(bh)} \\
M_\mathbf{k}^{(ag)} & 0 & M_\mathbf{k}^{(ae)} & 0\\
0 & M_\mathbf{k}^{(bh)} & 0 & M_\mathbf{k}^{(bf)} 
\end{pmatrix},
\end{gather*}
\begin{eqnarray*}
   M_\mathbf k^{(aa)} &=&  2J_{2b} S_1 \cos(k_x - k_y) + 2S_1 (J_{5a} + J_{5b}-J_{2a} - J_{2b}) ,  \\
  M_\mathbf k^{(bb)} &=& 2J_{2b} S_2 \cos(k_x - k_y) + 2S_2 ( J_{5a} + J_{5b}-J_{2a} - J_{2b}) ,\\
   M_\mathbf k^{(ac)} &=& -2J_{2a} S_1  \cos(k_x + k_y), \\
    M_\mathbf k^{(bd)} &=& -2J_{2a} S_2  \cos(k_x + k_y), \\
    M_\mathbf k^{(ae)} &=&  2J_{5a}S_1 \cos(k_x - k_y) , \\
     M_\mathbf k^{(bf)} &=&  2J_{5a}S_2 \cos(k_x - k_y) , \\
    M_\mathbf k^{(ag)} &=& -2J_{5b} S_1 \cos(k_x + k_y), \\
    M_\mathbf k^{(bh)} &=& -2J_{5b} S_2 \cos(k_x + k_y).
\end{eqnarray*}
For the \ensuremath{A_{1g}}\ ($LL$) channel, $M_\mathbf{k}=\overline K_\mathbf{k}/2$ where $\overline K_\mathbf{k}$ is the same dynamical matrix  $K_\mathbf{k}$ from Eq.~(\ref{eq4}) but with $J_c$ set to zero.

To calculate the Raman response, we transform $\mathcal{H}_\mathrm{FL}$ into the magnon basis $\Phi'_\mathbf{k}$ using the Bogoliubov matrix $T_\mathbf{k}$ derived previously: 
\begin{equation} \label{eq6} 
\mathcal{H}_\mathrm{FL}=\sum_\mathbf{k}\Phi_\mathbf{k}^\dag M_\mathbf{k} \Phi_\mathbf{k}
=\sum_\mathbf{k}{\Phi'}^\dag_\mathbf{k} M'_\mathbf{k} \Phi'_\mathbf{k},
\end{equation}
where $M'_\mathbf{k}=T_\mathbf{k}^\dag M_\mathbf{k}T_\mathbf{k}$ is the Raman vertex in the magnon eigenbasis. The final expression Eq.~(\ref{eq6}) contains the density-like terms, the substitutional terms, and the 2M terms. A density-like term (e.g., $\alpha_{1,\mathbf k}^\dag\alpha_{1,\mathbf k}$) generates an elastic response, and a substitutional term (e.g., $\alpha_{1,\mathbf k}^\dag\alpha_{2,\mathbf k}$) replaces a thermally populated magnon with another magnon at same $\mathbf k$, which is suppressed exponentially $\sim\mathrm{exp}(-E_{2,\mathbf k}/k_\mathrm{B}T)$ at low temperatures. 
The 2M terms (e.g., $\alpha_{1,\mathbf k}^\dag\alpha_{5,-\mathbf k}^\dag$) produce the dominant inelastic process at low temperatures. Therefore, only the 2M terms are retained in calculating the Raman response below.

The non-interacting 2M Raman spectrum ($\Im \chi_0$) is calculated as a retarded linear response function expressed with the transformed matrix elements, 
$M'_{ij}(\mathbf k)=\mel*{\alpha_{i,\mathbf k}^\dag\alpha_{j,-\mathbf k}^\dag}{\mathcal{H}_\mathrm{FL}}{0}$:
\begin{equation}\label{eq7} \begin{split}
\chi_0(\omega) =& \frac{1}{N}\sum_{\mathbf k,i,j}\abs{M'_{ij}(\mathbf k)}^2
\left(1-e^{-E_{ij}(\mathbf k)/k_\mathrm BT}\right)\\ &\times
\left(\frac{1}{\hslash\omega-E_{ij}(\mathbf k)+i\Gamma}
-\frac{1}{\hslash\omega+E_{ij}(\mathbf k)+i\Gamma}\right).
\end{split}\end{equation}
Here, the sum is over all $\mathbf k$-points in the Brillouin zone and all pairs of magnon branches $(i,j)$. $E_{ij}(\mathbf k)=E_i(\mathbf k)+E_j(\mathbf k)$ is the corresponding total 2M pair energy. For the numerical calculation, the summation was performed over a $2000\times2000$ $\mathbf k$-point mesh with a constant Lorentzian broadening of $\Gamma\equiv1$~meV.

To account for magnon-magnon interactions $J_\alpha$, we first define a reduced bare susceptibility 
$\overline\chi_0=\chi_0/\overline M$, which represents the intrinsic lineshape of the response, normalized by the mean-squared Raman vertex $\overline M=\frac{1}{N}\sum_{\mathbf{k},i,j}\abs*{M'_{ij}(\mathbf k)}^2$. 
For the simple antiferromagnets, $\overline\chi_0$ is equivalent to the bare susceptibility calculated using the well-known dimensionless geometric form factors, such as  $(\cos k_x - \cos k_y)^2$ for the \ensuremath{B_{1g}}\ channel \cite{brya74prb, elliott69jpc}. The final interacting Raman response is then obtained via the RPA-type renormalization,
\begin{equation}	\label{eq8}	
\Im \chi_\textrm{int} = \Im \frac{\chi_0}{1+J_\alpha\cdot\overline\chi_0}.
\end{equation}
To compare the final theoretical spectra with the experimental results, contributions arising from the high- and low-spin moments $m_i$ ($i=1,2$)  were scaled separately each with a single, common multiplication factor ($\mathcal{A}_i/S_i^2$) applied to both the \ensuremath{B_{1g}}\ and \ensuremath{B_{2g}}\ channels. Supplementary calculations of the magnon dispersions and Raman spectra, obtained using parameters differing from the values in Table~\ref{tab1}, are presented in Figs.~\ref{fig-15} and \ref{fig-16}.

\section{Reproducibility} \label{app:f}

\begin{figure}[b] 		
{\includegraphics[width=1.0\columnwidth,draft=false]{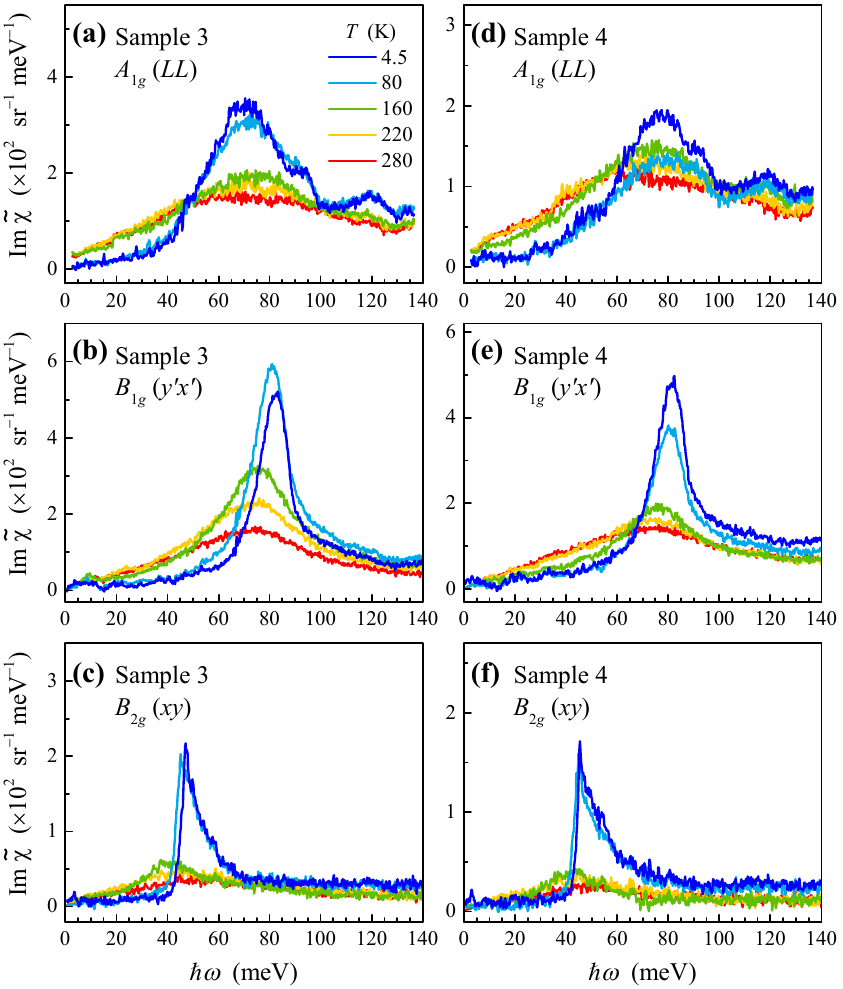}} \vspace{-2.0em}
\caption{\label{fig-17}%
{Polarized Raman responses $\Im\tilde\chi$ of (a)-(c) Sample~3 and (d)-(f) Sample~4 at five representative temperatures (4.5, 80, 160, 220, 280~K), measured within the \ensuremath{A_{1g}}, \ensuremath{B_{1g}}, and \ensuremath{B_{2g}}\ scattering channels, respectively.}
}
\end{figure}

\begin{figure}[] 		
{\includegraphics[width=0.85\columnwidth,draft=false]{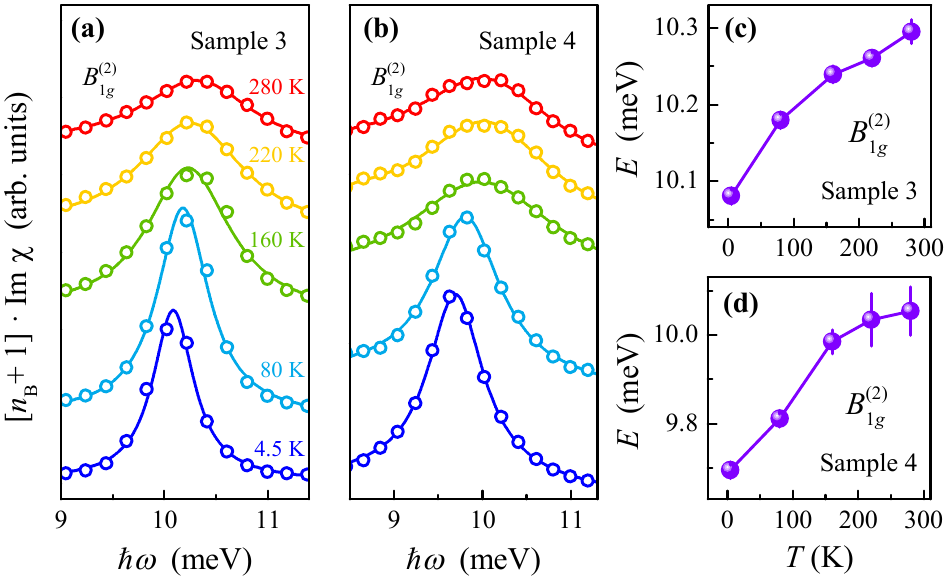}} \vspace{-1.0em}
\caption{\label{fig-18}%
{(a),(b) Soft $B_{1g}^{(2)}$ phonon mode spectra of Samples~3 and 4 at each temperature, plotted with vertical offsets.
(c),(d) Temperature-dependent $B_{1g}^{(2)}$ phonon energies of Samples~3 and 4.}
}
\end{figure}

It is essential to confirm whether the main findings of the present study are reproducible across different \ce{La3Ni2O7}\ crystal pieces.
The main temperature-dependent Raman data analyzed in the main text are obtained from Sample~1 [Fig.~\ref{fig5}(d1)].
To check the reproducibility, additional temperature-dependent Raman experiments are performed on new crystal pieces, Samples~3 and 4 [Figs.~\ref{fig5}(d3),(d4)].

As summarized in Figs.~\ref{fig-17} and \ref{fig-18}, the central findings of our study are robustly reproduced in these new crystals:
(i) In the \ensuremath{A_{1g}}\ channel, the low-energy electronic continuum below 50~meV is abruptly depleted below 160~K, becoming nearly zero below 80~K [Figs.~\ref{fig-17}(a),(d)];
(ii) The \ensuremath{B_{1g}}\ and \ensuremath{B_{2g}}\ channels feature prominent 2M excitation signals [Figs.~\ref{fig-17}(b),(c),(e),(f)] with lineshapes consistent with those in the main text;  
(iii) The low-energy \ensuremath{B_{1g}}\ phonons soften continuously from 280~K down to 4.5~K (Fig.~\ref{fig-18}), corroborating the presence of an incipient lattice softness of the \ensuremath{B_{1g}}\ symmetry across different \ce{La3Ni2O7}\ crystals.

It is noted that minor variances are identified in the Raman spectra of different crystal pieces.
For instance, the 2M peak energies vary by $<2$~meV (Fig.~\ref{fig-17}) and phonon energies show slight variations of a few percent depending on the specific sample (Fig.~\ref{fig-18}).
Signal intensities also depend on the condition of the cleaved surface---for example, the broad \ensuremath{A_{1g}}\ hump around 70~meV of Sample~4 is slightly weaker than in other samples at low temperatures [Fig.~\ref{fig-17}(d)].
Quantifying the connection between these variances and the sample quality can be an important avenue for future study.
However, the primary focus of the present study is to investigate the intrinsic and universal properties of the bilayer nickelate, rather than detailing sample-to-sample variations.
Because the characteristic thermal evolution of the central Raman signatures is consistently reproduced across multiple specimens, we conclude that the phenomena reported in the main text  reflect the true nature of the SDW order and faithfully represent the intrinsic characteristics of \ce{La3Ni2O7}\ at ambient pressure, which are not sample-dependent artifacts.

\bibliography{ref}
\end{document}